\documentclass[a4paper,11pt]{article}
\usepackage{jheppub}

\usepackage[T1]{fontenc} 

\usepackage{chessfss}
\pdfoutput=1
\pdfpageattr {/Group << /S /Transparency /I true /CS /DeviceRGB>>}
\hypersetup{colorlinks=true,linkcolor=blue,citecolor=blue}
\usepackage{amsmath,amssymb,amsfonts,mathrsfs}
\usepackage{bm}
\usepackage{float}
\usepackage{subfig}
\usepackage{mathtools}
\usepackage{scalefnt}
\usepackage{wrapfig}
\usepackage{todonotes}
\usepackage{fancybox}
\usepackage{graphicx}
\usepackage{tikz-cd}
\usepackage{multirow}
\usepackage{adjustbox}
\usepackage[toc,page]{appendix}

\allowdisplaybreaks

\def\CA{{\cal A}}

\def\CC{{\cal C}}

\def\CE{{\cal E}}

\def\CH{{\cal H}}
\def\CI{{\cal I}}

\def\CR{{\cal R}}
\def\CM{{\cal M}}
\def\CN{{\cal N}}
\def\CO{{\cal O}}

\def\beq#1\eeq{\begin{align}#1\end{align}}

\newcommand{\be}{\begin{eqnarray}}
\newcommand{\ee}{\end{eqnarray}}
\newcommand{\bea}{\begin{eqnarray}}
\newcommand{\eea}{\end{eqnarray}}

\newcommand\gmod{\operatorname{\frak{g}-\mathrm{mod}}}

\newcommand{\bn}{\begin{enumerate}}
\newcommand{\en}{\end{enumerate}}

\def\identity{{\rlap{1} \hskip 1.6pt \hbox{1}}}
\def\iden{\identity}

\def\IC{\mathbb{C}}

\def\IP{\mathbb{P}}

\def\IR{\mathbb{R}}
\def\IZ{\mathbb{Z}}

\def\CA{{\cal A}}

\def\CC{{\cal C}}

\def\CE{{\cal E}}
\def\CF{{\cal F}}

\def\CH{{\cal H}}
\def\CI{{\cal I}}

\def\CM{{\cal M}}
\def\CN{{\cal N}}
\def\CO{{\cal O}}

\def\CR{{\cal R}}

\def\half{\frac{1}{2}}

\def\det{{\rm det}}

\usepackage{mathtools,amsmath}
\usepackage{verbatim}

\title{Foams and KZ-equations in Rozansky-Witten theories}

\begin{document}

\author[a,b]{Sergei Gukov,}
\author[c,d]{Babak Haghighat,}
\author[c,d]{Nicolai Reshetikhin}

\affiliation[a]{Richard N. Merkin Center for Pure and Applied Mathematics, California Institute of Technology, Pasadena, CA 91125, USA}
\affiliation[b]{Dublin Institute for Advanced Studies, 10 Burlington Rd, Dublin, Ireland}
\affiliation[c]{Yau Mathematical Sciences Center, Tsinghua University, Beijing, 100084, China}
\affiliation[d]{Beijing Institute of Mathematical Sciences and Applications (BIMSA), Huairou District, Beijing 101408, China}

\abstract{In this paper, we present a geometric description of foams, which are prevalent in topological quantum field theories (TQFTs) based on quantum algebra, and reciprocally explore the geometry of Rozansky-Witten (RW) theory from an algebraic perspective. This approach illuminates various aspects of decorated TQFTs via geometry of the target space $X$ of RW theory. Through the formulation of the Knizhnik-Zamolodchikov (KZ) equation within this geometric framework, we derive the corresponding braiding and associator morphisms. We discuss applications where the target space of RW theory emerges as the Coulomb branch of a compactified 6d SCFT or Little String Theory, with the latter being particularly intriguing as it results in a compact $X$.}
%

\maketitle

\section{Introduction}

Our goal in this paper is to tie together several topics that all fit under the umbrella of topological quantum field theory (TQFT) but so far were studied independently and had little contact with one another. These include the emerging significance of foams in knot homologies, the rapidly evolving field of generalized (categorical) symmetries, the increasing relevance of ``decorated TQFTs,'' and numerous connections with symplectic and algebraic geometry.

A natural context where these diverse perspectives converge is within Rozansky-Witten theories \cite{Rozansky:1996bq,MR1671725,Kapranov}. Each such theory is defined by selecting a complex symplectic space $X$, which may be singular or non-compact, or both. By focusing on this class of TQFTs, we inherently establish connections to geometry. By identifying and exploring structures commonly associated with quantum algebra, we effectively construct a bridge between algebra and geometry,
$$
\textbf{Geometry} \quad \longleftrightarrow \quad \textbf{Algebra}
$$
This allows us to express known algebraic structures of 3d TQFTs in terms of the geometry of an appropriate target space $X$, and vice versa. A pertinent example is the Knizhnik-Zamolodchikov (KZ) equation, which plays a crucial role in TQFTs derived from algebraic inputs like quantum groups or quantum algebras via the Reshetikhin-Turaev construction \cite{reshetikhin1991invariants}. One of our aims is to examine the braiding of line operators in a Rozansky-Witten theory with target space $X$ and to elucidate the ``geometric'' analogues of the KZ equation.

Another impetus for this work is to bridge two parallel lines of development, each offering complementary views on ``decorated TQFTs'' --- topological quantum field theories defined on manifolds with additional data (background fields).\footnote{See e.g. \cite{Jagadale:2022abr} for a recent overview and applications in context relevant to this work.} One approach is rooted in generalized symmetries \cite{Gaiotto:2014kfa}, emphasizing topological operators supported on submanifolds of various (co-)dimensions that can form junctions and interact. The other, emerging from the development of homological knot invariants, refers to these junctions as ``foams,'' see e.g. \cite{MR3880205,MR4164001,MR3877770} for a small sample of recent developments.

In 3d TQFTs, both lines of development involve junctions of topological operators supported on surfaces (a.k.a. surface operators). It is natural to investigate whether a connection exists between the two. We argue that, at least within the context of Rozansky-Witten theories, there is indeed a close relationship. Specifically, we offer a geometric framework for classifying foams based on the geometry and topology of $X$, in a way that is governed by the generalized symmetries of the theory. The study of topological defect operators supported on codimension one and higher surfaces was initiated in \cite{Gaiotto:2014kfa} and has been a very active field since, see references \cite{McGreevy:2022oyu,Cordova:2022ruw,Brennan:2023mmt,Bhardwaj:2023kri,Schafer-Nameki:2023jdn,Luo:2023ive,Shao:2023gho,Carqueville:2023jhb} for comprehensive reviews. In particular, in the context of 3d QFTs (but so far not topological), non-invertible surface defects have been constructed by performing gauging in half spacetime \cite{Choi:2024rjm,Cui:2024cav}. Our more geometric approach to the construction of such defects in the context of the Rozansky-Witten TQFT can complement these works by providing a new alternative perspective.

We anticipate that this study will have various applications and extensions, particularly to other types of ``decorated TQFTs'' and aspects of geometric representation theory where quantum groups are significant. Given that Coulomb branches of supersymmetric gauge theories with a Lagrangian description are always non-compact, resulting in Rozansky-Witten TQFTs that are either non-unitary \cite{Dedushenko:2018bpp} or non-semisimple \cite{Gukov:2020lqm}, we hope the insights gained from this work will aid in the future understanding of such TQFTs. In contrast, Coulomb branches of compactified Little String Theories can be compact \cite{Intriligator_2000}, and we explore a detailed example in section~\ref{sec:LST}.

The organization of this paper is as follows. In Section \ref{sec:symmetries} we present an overview of the categoriacal symmetries of Rozansky-Witten theory. In Section \ref{sec:MTC} we review the notion of a modular tensor category. Section \ref{sec:RWtheory} represents the main body of the paper, where we define the braiding and associator morphisms for Rozansky-Witten theory, show that they satisfy the correct properties and explore their action on the Hilbert space. Here we also identify invertible and non-invertible one-form and zero-form symmetries. In Section \ref{sec:LST} we show how Rozansky-Witten theory with a K3 target can be obtained from Little String Theory compactifications on $T^3$ and explore the properties of defect operators in such theories. Finally, in Section \ref{sec:applications} we present applications of the some of the ideas developed.

\section{Categorical symmetries in Rozansky-Witten theory}
\label{sec:symmetries}

This section presents an overview of the cagegorical structure of Rozansky-Witten theory and highlights some of the main lessons drawn from the subsequent sections.

\subsection{Basics}

Rozansky and Witten introduced a 3d topological field theory formulated as a sigma model with target space a hyperk\"ahler manifold \cite{Rozansky:1996bq}. Let $X$ be a hyper-K\"ahler manifold of real dimension $4n$. The complexification of the tangent bundle admits a decomposition
\begin{equation}
    TX \otimes_{\IR} \IC = V \otimes S,
\end{equation}
where $V$ is a rank $2n$ complex vector bundle with structure group $Sp(n)$, and $S$ is a trivial rank two bundle. Denote local coordinates on a $3$-manifold $M$ as $x^{\mu}$, $\mu=1,2,3$. Define a TQFT with fields:
\begin{itemize}
    \item Bosons: $\Phi: M \rightarrow X$, $\phi^i(x^{\mu})$, $i=1,\ldots,4n$
    \item Fermions: Scalar fields $\eta^I$ and one-forms $\chi_{\mu}^I$ with values in $V$
\end{itemize}
and action given by ($\Omega$ is a completely symmetric tensor),
\begin{eqnarray}
    S & = & \int_M (L_1 + L_2) \sqrt{h}d^3x, \nonumber \\
    L_1 & = & \half g_{ij}\partial_{\mu} \phi^i \partial^{\mu}\phi^j + \epsilon_{IJ} \chi^I_{\mu} \nabla^{\mu}\eta^J, \nonumber \\
    L_2 &  = & \half \frac{1}{\sqrt{h}}\epsilon^{\mu\nu\rho}\left(\epsilon_{IJ} \chi^I_{\mu} \nabla_{\nu} \chi^J_{\rho} + \frac{1}{3}\Omega_{IJKL}\chi^I_{\mu} \chi^J_{\nu} \chi^K_{\rho} \eta^L\right).
\end{eqnarray}

Roughly speaking, one can introduce BPS line defects in RW-theory by choosing a holomorphic vector bundle on the target manifold $X$ and coupling the action to the connection of this bundle. This procedure will be explained in more detail in Section \ref{sec:RWtheory}. For now it will suffice to note that the particular choice of a holomorphic bundle can be viewed as an object in the derived category of coherent sheaves on $X$, denoted by $D^b(X)$. As will be explained below, line defects labeled by objects in $D^b(X)$ can end on co-dimension one surface operators which act as autoequivalences $\mathrm{Auteq}D^b(X)$ on the line defects. 

\subsection{Categorical symmetries}

This section combines the analysis of generalized symmetries and the classification of foams in a general Rozansky-Witten theory with target space $X$. In fact, these two problems are simply the two sides of the same coin, as Table~\ref{tablefoams} illustrates.

\begin{table}[htb]
\centering
\begin{tabular}{ccc|c|c}
Role in topology: & \qquad &  & {\bf Webs} & {\bf Foams} \\
\cline{3-5}
\multirow{2}{*}{Topological operator:} & \qquad & ``local'' & line & surface \\
 & & operators & operators & operators \\
\cline{3-5}
Support in $M_3$: & & \phantom{$\oint^{\oint}_{\oint}$} points \phantom{$\oint^{\oint}_{\oint}$} & 1-dimensional & 2-dimensional \\
\cline{3-5}
Corresponding symmetry: & & 2-form & 1-form & 0-form \\
\cline{3-5}
Algebraic structure: & & $\CR$ & $\CA$ & $G$ \\
\end{tabular}
\caption{Types of topological operators in 3d, the corresponding generalized symmetries, and their role in low-dimensional topology.}
\label{tablefoams}
\end{table}

Indeed, on the one hand, in three dimensions topological operators of dimension $2-p$ implement $p$-form symmetries~\cite{Gaiotto:2014kfa,Benini:2018reh}. On the other hand, junctions of such operators encode the algebraic structure of the theory that describes webs (resp. foams) when $p=1$ (resp. $p=0$). Therefore, from either perspective the problem boils down to the classification of topological operators of various dimension, and we introduce the following notations for the algebraic structures that classify their types (``charges''):
\begin{itemize}

\item $G:$ 0-form symmetry, foams

\item $\CA:$ 1-form symmetry, lines

\item $\CR:$ 2-form symmetry, local operators

\end{itemize}
\noindent
One of our goals in this section is to describe each of these sets geometrically, in terms of~$X$.

Below we discuss each one in turn, and then consider how they interact with each other.

\begin{figure}[ht]
	\centering
	\includegraphics[trim={0.2in 0 0.4in 0},clip,width=2.6in]{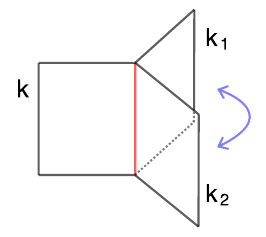}
	\caption{A junction (seam) of surface operators in 3d.}
	\label{fig:junction1}
\end{figure}

\subsection{Foams}

Let us start with surface operators, i.e. 2-dimensional topological interfaces. The fusion (``stacking'') of such topological defects defines a group law, which does not need to be commutative since we are considering codimension-1 operators that ``do not have enough room'' to go around one another. We denote the resulting group by $G$.

\begin{figure}[ht]
	\centering
	\includegraphics[trim={1.1in 0.2in 1.1in 0.8in},clip,width=2.6in]{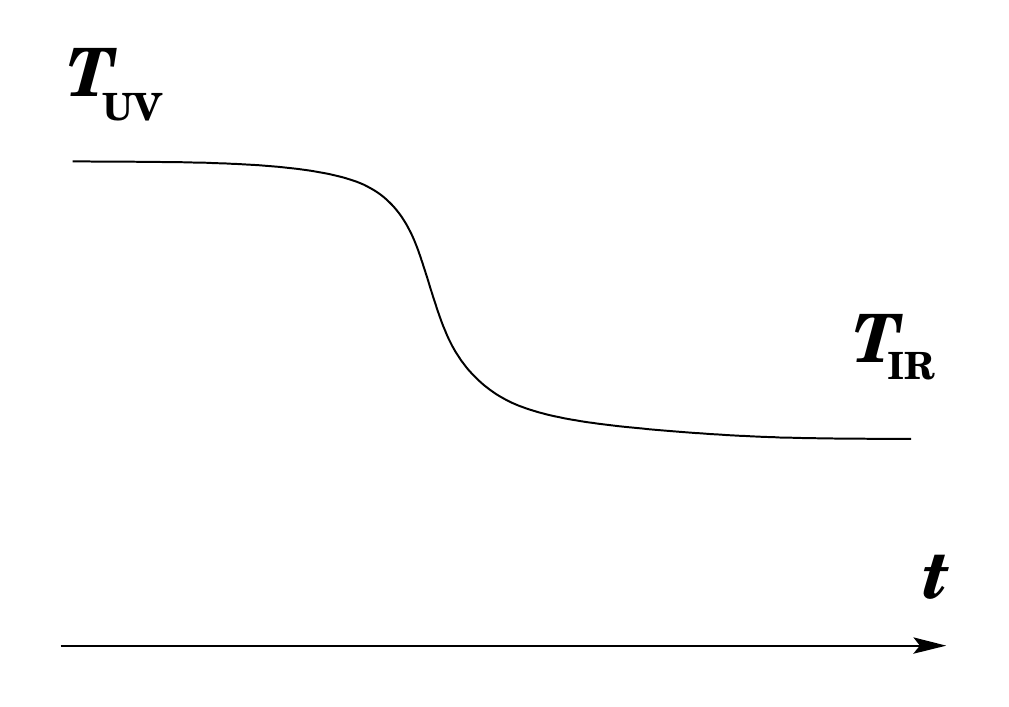}
	\caption{Codimension-1 surface operators / walls / interfaces can be engineered by varying parameters of the theory in one spatial dimension.}
	\label{fig:parameters}
\end{figure}

In topological sigma-models, one way to find the spectrum of codimension-1 topological operators is to consider the space of parameters of the underlying physical theory which become $Q$-exact in the topologically twisted theory. Then, an adiabatic change of such parameters with one of the space-time dimensions, say $x^1$, defines a codimension-1 object (a wall, or interface), illustrated in Figure~\ref{fig:parameters}. Because the theory is topological, the thickness of such a domain wall can be changed at will, in particular it can be made arbitrarily small, making such monodromy defects fully localized in one of the space-time dimensions. This change can be thought of as the change of metric components on the underlying 3-manifold. And, if the variation of the $Q$-exact parameters traverses a closed loop in the space of parameters, it means that the wall / interface has the same TQFT on both sides. The set of all possible monodromies is clearly parametrized by the fundamental group of the space of $Q$-exact parameters (stability conditions), so that we conclude \cite{Gukov:2006jk} (see also \cite{Gadde:2013wq}):
\be
G \; = \; \pi_1 \left( \{ \text{parameters} \} \right)
\label{Gparam}
\ee
This reasoning was used to provide a geometric and gauge theoretic realization of the categorified affine Hecke algebra as algebra of interfaces \cite{Gukov:2006jk}. The case of 3d Rozansky-Witten theory is very similar.

In Rozansky-Witten theory, the underlying physical theory is a 3d $\CN=4$ sigma-model with target space $X$. The holomorphic symplectic structure on $X$, say, in complex structure $J$, is part of the definition of the Rozansky-Witten theory. However, the complexified K\"ahler structure associated with $J$ is $Q$-exact. (It can be viewed as a choice of stability conditions.) Monodromies in the space of such K\"ahler structures (stability conditions) are expected to be precisely the autoequivalences of the derived category of coherent shaves,
\be
D^b (X) := D^b \text{Coh} (X)
\label{DbX}
\ee
in complex structure $J$. Therefore, we expect
\be
G \; = \; \text{Auteq} \, D^b (X)
\label{Gauteq}
\ee
Below we provide further justification of this conclusion.

Before we turn to the next case of line operators, let us point out that the group law (product operation) in $G$ is precisely what makes it possible to construct foams from surface operators. Indeed, foams have seams (junctions) where three surface operators labeled by $g_1$, $g_2$, and $g_1 g_2$ come together, as illustrated in Figure~\ref{fig:junction1}.

\begin{figure}[ht]
	\centering
	\includegraphics[trim={0 0 0 0},clip,width=2.6in]{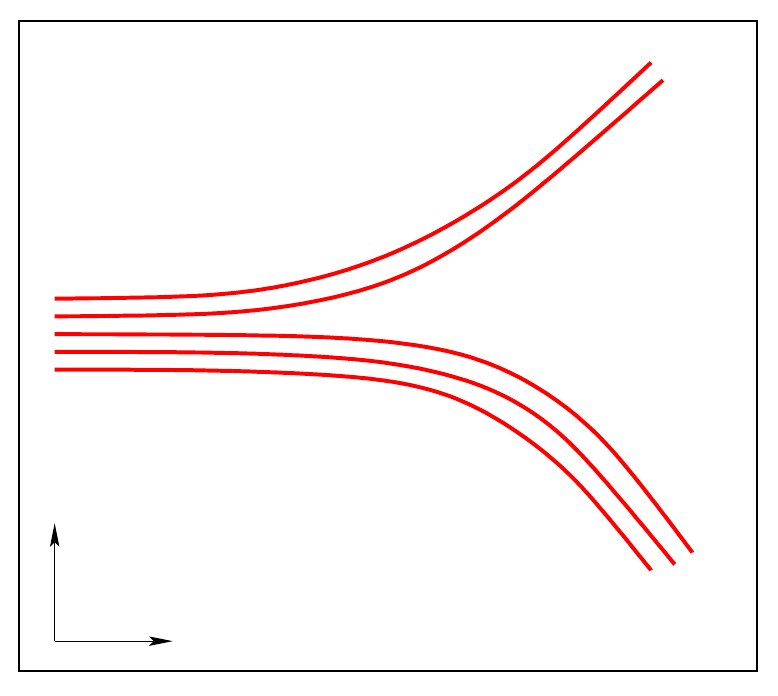}
	\caption{In string theory, junctions of surface operators sometimes can be realized as stacks of branes coming together. (Shown here is the ``view from the top'' perspective on Figure~\ref{fig:junction1}.)}
	\label{fig:junction4}
\end{figure}

\subsection{Lines}

Unlike surface operators, line operators in Rozansky-Witten are much better studied. In fact, in the original work of Rozansky and Witten \cite{Rozansky:1996bq} it is explained how one can associate a line operator to a holomorphic bundle on $X$. Since cohomological TQFTs typically lead to a derived setting, it is natural to expect that line operators correspond to objects in the derived category of coherent sheaves on $X$. And, indeed, it was advocated in \cite{MR2661534} that (a 2-periodic version of) the category \eqref{DbX} classifies line operators. We give it a name
\be
\CC := \text{line opertors} = D^b (X)
\label{linecategory}
\ee

Before we proceed, let us point out that in the opening of this section we introduced an algebraic structure $\CA$ that classifies types (``charges'') of line operators in our three-dimensional topological theory. One may say that $\CA$ classifies $1$-form symmetries. The charges take values in the Grothendieck group of the corresponding category and, therefore, we have the following relation
\be
\CA = K^0 (\CC)
\label{linecharges}
\ee
For a hyper-K\"ahler manifold and the category \eqref{DbX}, this is isomorphic to the cohomology of $X$, which agrees with the result of Rozansky and Witten \cite{Rozansky:1996bq} if we combine it with the state-operator correspondence:
\be
\CH_{\text{TQFT}} (T^2) = \left[\text{line opertors}\right] = K^0 (\CC)
\label{torusspace}
\ee
Two remarks are in order. First, this relation is very general and not limited to Rozansky-Witten theories. Second, in the relation \eqref{torusspace}, ``line operators'' means only classification of charges of line operators, not to be confused with a more refined classification given by \eqref{linecategory}, which is indicated here by the use of brackets.

Much like surface operators can have junctions that lead to foams, line operators can form junctions that lead to webs. We shall discuss such configurations shortly, after we review the spectrum of local operators in Rozansky-Witten theory with target space $X$.

Meanwhile, it is instructive to see that the answers we have obtained for $G$, $\CC$, and $\CA$ are all compatible with each other.
Indeed, in a general 3d QFT --- not necessarily topological! --- when a topological line operator of type $a \in \CA$ goes through a topological surface operator (foam) labeled by $g \in G$, it transforms into a line operator $\rho_g a$, where
\be
\rho: G \to \text{Aut} (\CA)
\ee
In a general 3d QFT, the data of $G$ and $\CA$ encodes the algebraic structure of 0- and 1-form symmetries, respectively. In this more general context, the choice of $\rho$ is an extra data that enters the definition of theory and is part of what is called a 2-group structure~\cite{Benini:2018reh}.

In our setup, the entire 3d theory is topological, and the definition of the Rozansky-Witten theory does not seem to involve any additional choices once we fix the target space $X$. This suggests that there is a canonical choice of $\rho$ and, indeed, it is precisely what our answers for $G$, $\CC$, and $\CA$ tell us, cf. \eqref{Gauteq}, \eqref{linecategory}, and \eqref{linecharges}.

Another potential ``mixing'' --- or, more precisely, 2-group structure --- between $G$ and $\CA$ may involve a choice of the Postnikov class
\be
[\beta] \in H^3_{\rho} (BG, \CA)
\ee
or, equivalently,
\be
\beta: G \times G \times G \to \CA
\ee
This class describes a junction (a version of the ``F-move'' for foams) where four surface operators --- labeled by $g_1$, $g_2$, $g_3$, and $g_2 g_2 g_3$ --- meet and one line operator of type $\beta (g_1,g_2,g_3)$ emanates from the common intersection point.
Just like with the choice of $\rho$, one might expect {\it a priori} that in Rozansky-Witten theory $\beta$ has a canonical value. Indeed, since the Postnikov class $\beta$ can be interpreted as non-associativity of surface operators (foams) acting on lines, in Rozansky-Witten theory it vanishes
\be
\beta = 0
\ee

\subsection{Local operators}

It remains to describe the spectrum of local operators, $\CR$, and the role it plays for line and surface operators. Similarly to \eqref{torusspace}, by state-operator correspondence we have
\be
\CH_{\text{TQFT}} (S^2) = \text{local opertors} =: \CR
\label{spherespace}
\ee
Then, using the result of \cite{Rozansky:1996bq} for the left-hand side, we can identify $\CR$ with the commutative ring:
\be
\CR = \mathbb{C} [X]
\label{RviaX}
\ee
This ring is commutative because the product (defined by fusing local operators) has no order; local, point-like operators in 3d can easily go around one another to switch the order.

\begin{figure}[ht]
	\centering
	\includegraphics[clip, trim=1cm 6cm 1cm 6cm, width=0.60\textwidth]{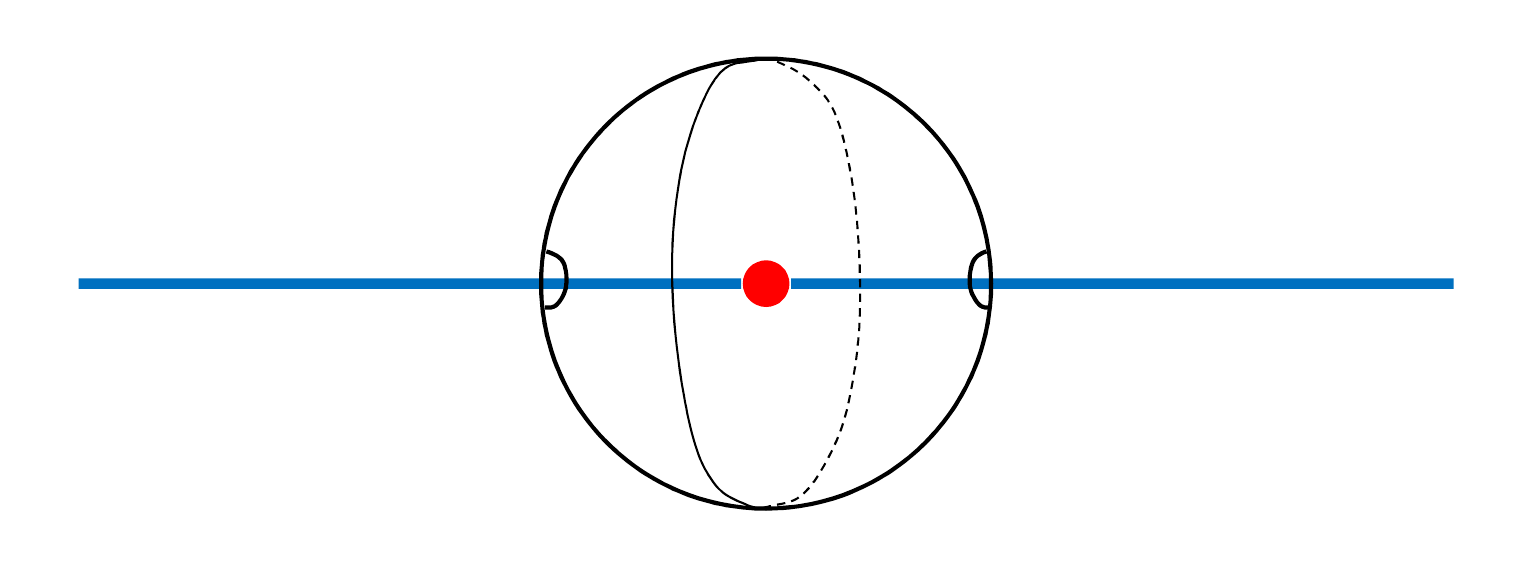}
	\caption{Two topological line operators, $\CE$ and $\CF$, can meet at a point that supports a local operator $x \in \text{Ext} (\CE, \CF)$.}
	\label{fig:linejunction}
\end{figure}

Now we are ready to discuss junctions of line operators that was promised earlier and, in particular, to see that \eqref{RviaX} is consistent with the algebraic structure of line operators. The category \eqref{linecategory} has a monoidal structure, such that the unit object $\bf{1} \in \CC$ corresponds to the structure sheaf $\CO_X$ (see {\it e.g.} \cite{Qiu:2020mji}). Moreover, the spectrum of local (point-like) operators at the junction of two line operators $\CE$ and $\CF$ is given by
$$
\text{Ext} (\CE, \CF)
$$
In particular, if we take both $\CE$ and $\CF$ to be unit objects, $\bf{1} \in \CC$, then we recover the spectrum of local operators supported at a point in $M_3$:
$$
\text{Ext} ({\bf 1}, {\bf 1}) = \CO_X = \CR
$$
Indeed, the trivial object in the category $\CC$ corresponds to no line operator at all, and in this case the Figure~\ref{fig:linejunction} illustrates the spectrum of local topological operators surrounded by an ordinary (not punctured) 2-dimensional sphere.

\section{Modular Tensor Categories}
\label{sec:MTC}

As we have seen in the previous section, line operators of RW-theory are labeled by objects in the categery $\mathcal{C}=D^b(X)$. In a 3d topological QFT, such line operators often exhibit more structure including braiding and a non-trivial action on $\mathcal{H}_{\mathrm{TQFT}}(\partial M_3)$ under large diffeomorphisms of the boundary of $M_3$. This leads to the structure of a modular tensor category to which we turn next.

To define braiding in a topological field theory, we need a couple of ingredients which we shall define now where we will be mainly following the notation of \cite{RowellTensor,bakalov2001lectures}. Recall that a \textit{monoidal category} $(\mathcal{C},\otimes,\iden)$ is a category with a tensor product $\otimes$ and an identity object $\iden$, 
\begin{itemize}
    \item $\otimes$ is associative in a sense that for each triple $A, B, C \in \CC$ there exists a functorial isomorphism\footnote{Note that in the mathematical literature $F$ is usually denoted by $a$. In the physics literature one uses $F$, which might originate from the word ``fusion''.}
    \begin{equation}
        F_{A,B,C} : A \otimes (B \otimes C) \rightarrow (A \otimes B) \otimes C,
    \end{equation}
    that satisfies the pentagon identity, see for example \cite{bakalov2001lectures}. Recall that the pentagon identity is the commutativity of the diagram shown in Figure \ref{fig:Pentagon},
    \begin{figure}[h!]
    \centering
    \includegraphics[width=0.8\textwidth]{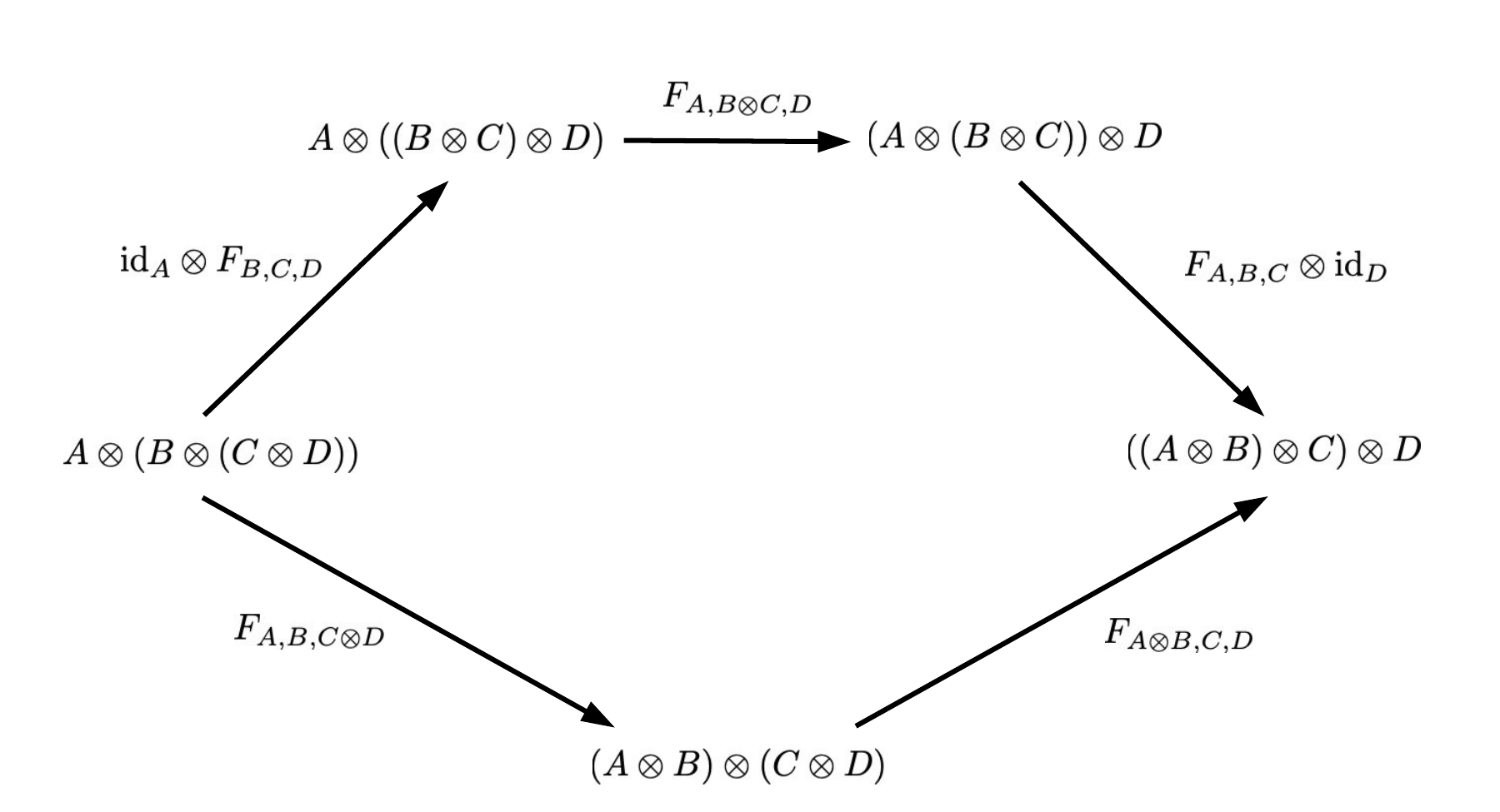}\caption{The pentagon identity}\label{fig:Pentagon}
    \end{figure}
    while the associativity morphisms $F_{A,B,C}$ can be represented by the diagram in Figure \ref{fig:Associator}.
    \begin{figure}[h!]
    \centering\includegraphics[width=0.4\textwidth]{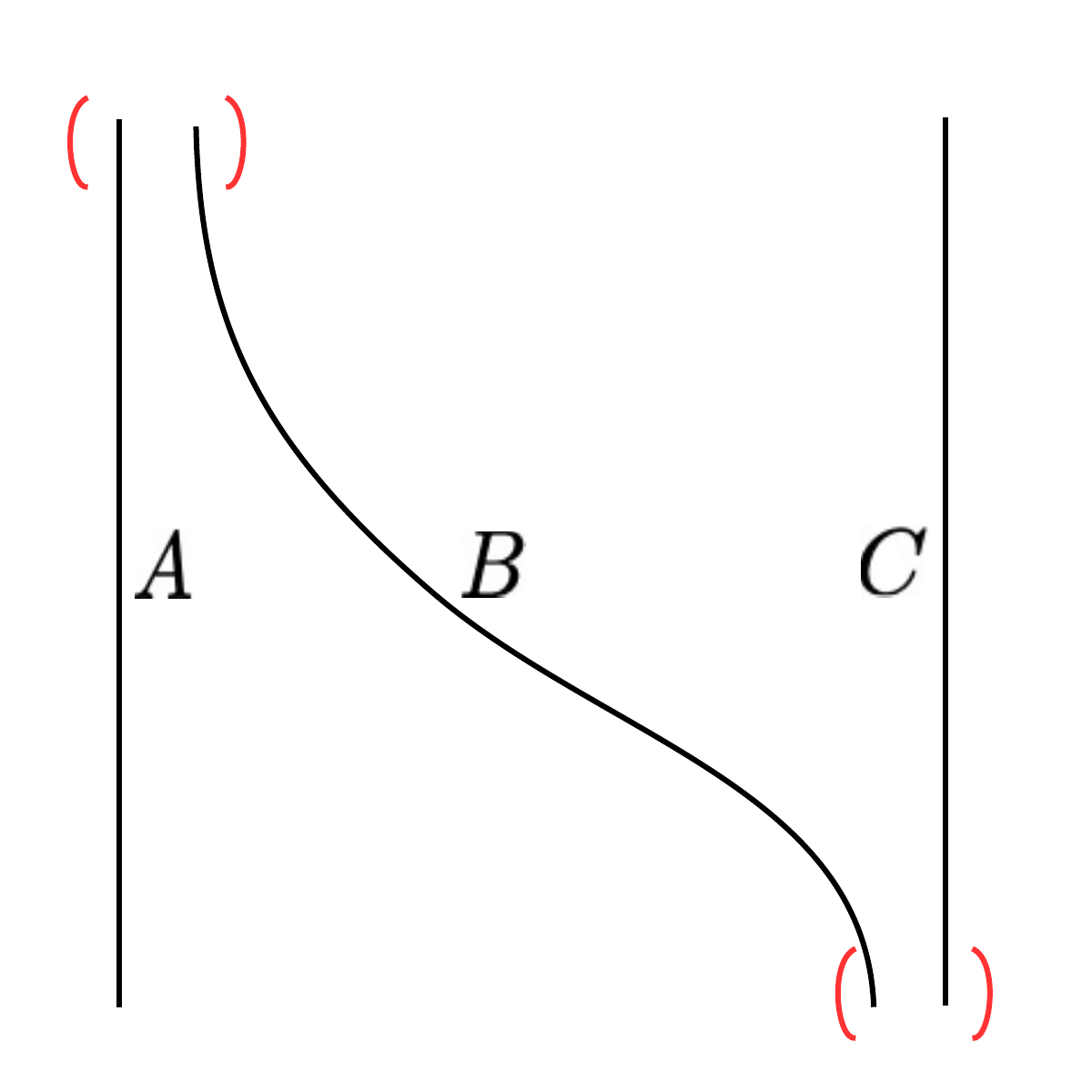}
  \caption{The associator}
  \label{fig:Associator}
\end{figure}
    Then the pentagon equation can be interpreted as the isotopy depicted in Figure \ref{fig:PentagonGraph}.
    \begin{figure}[h!]
    \centering\includegraphics[width=0.8\textwidth]{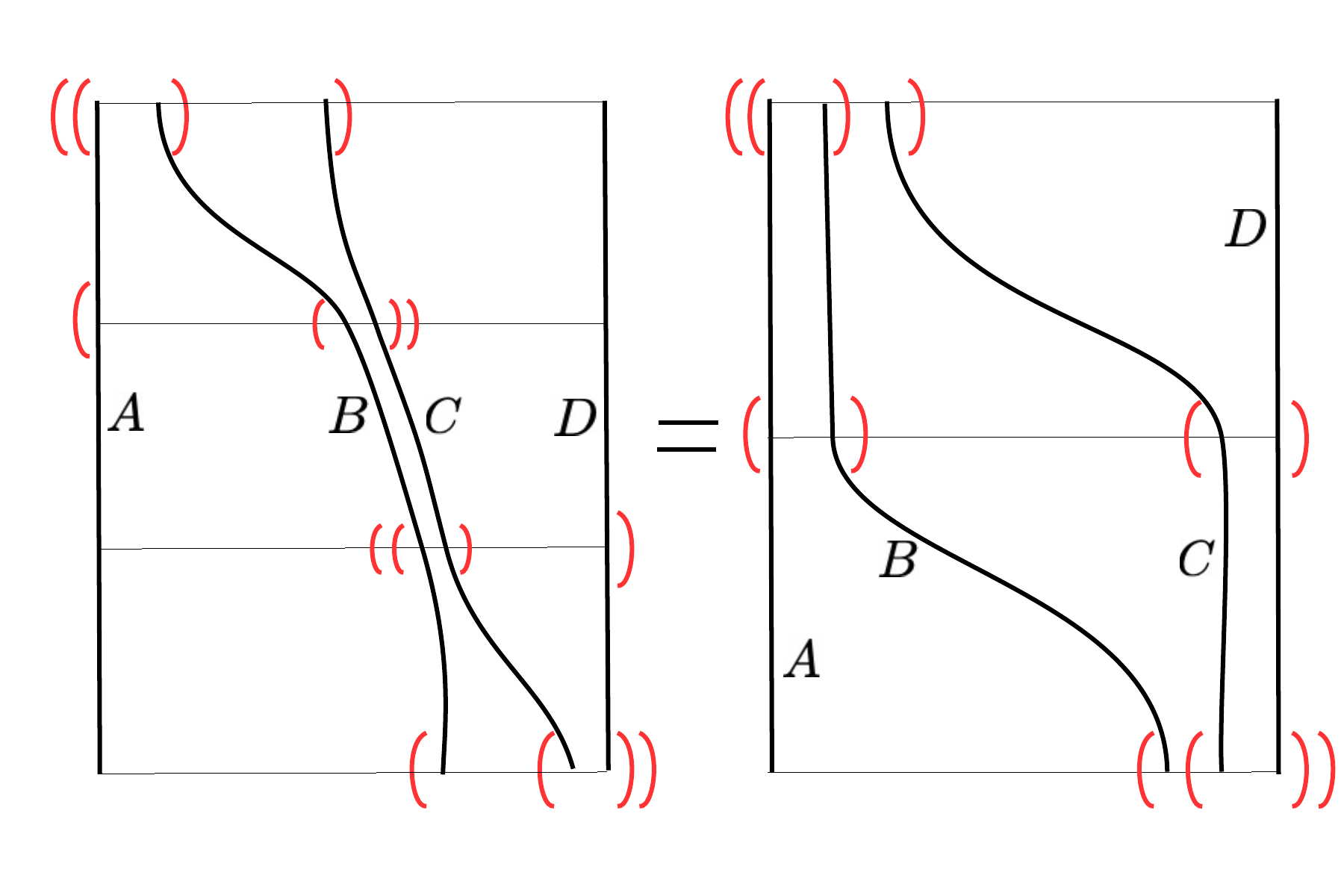}
  \caption{The pentagon isotopy}
  \label{fig:PentagonGraph}
\end{figure}
    \item for any $A \in \mathcal{C}$, we have natural isomorphisms
    \begin{equation}
        l_A : \iden \otimes A \rightarrow A, \quad r_A : A \otimes \iden \rightarrow A,
    \end{equation}
    that satisfy certain identities, see \cite{bakalov2001lectures} and references therein.
\end{itemize}

From now on we assume that $\CC$ is an Abelian category over an algebraically closed field $k$. For us $k$ will be always $\mathbb{C}$. Recall that $\CC$ is semisimple if every object is isomorphic to a direct sum of simple modules. We denote the subset of simple object by $\CI \subset \CC$. For a pair $A, B \in \CI$ we have
\begin{equation}
    \mathrm{Hom}(A,B) = \left\{\begin{array}{cc}
        k \iden_A,  & A = B \\
        0           & A \neq B     
    \end{array}\right.
\end{equation}
In particular, the identity object is simple.

A \textit{fusion category} is a semi-simple tensor category with finitely many simple objects. For convenience we enumerate simple objects as $\CI = \{A_i\}$. For simple objects, define the coefficient,
\begin{equation}
    N_k^{i,j} \equiv \mathrm{dim}(\mathrm{Hom}(A_k,A_i \otimes A_j)),
\end{equation}
as the multiplicity of the decomposition of $A_i \otimes A_j$ into single $A_k$. The data $\left\{N_k^{i,j}\right\}$ is the \textit{fusion ring} $[\CC]$ of the category $\mathcal{C}$. It consists of formal $\mathbb{Z}$-linear combinations of isomorphism classes of objects in $\CC$, and is generated by isomorphism classes of simple objects modulo relations
\begin{equation}
    [A_i]  \cdot [A_j] = \sum_k N_k^{i,j} [A_k].
\end{equation}
The category $\mathcal{C}$ is \textit{rigid} if any object $A \in \mathcal{C}$ has its' (right) \textit{dual} $A^*$, with the evaluation and coevaluation maps
\begin{eqnarray}
    \mathrm{ev}_A   & : & A^* \otimes A \rightarrow \iden, \nonumber \\
    \mathrm{ev}_A^* & : & \iden \rightarrow A^* \otimes A. 
\end{eqnarray}
\paragraph{Example.} 
The $MTC[M_3]$ described in Section \ref{sec:MTCM3} with fusion product given in equation \eqref{eq:latticefusion} is an example of such a fusion category.

\begin{figure}[h!]
\centering\includegraphics[width=0.8\textwidth]{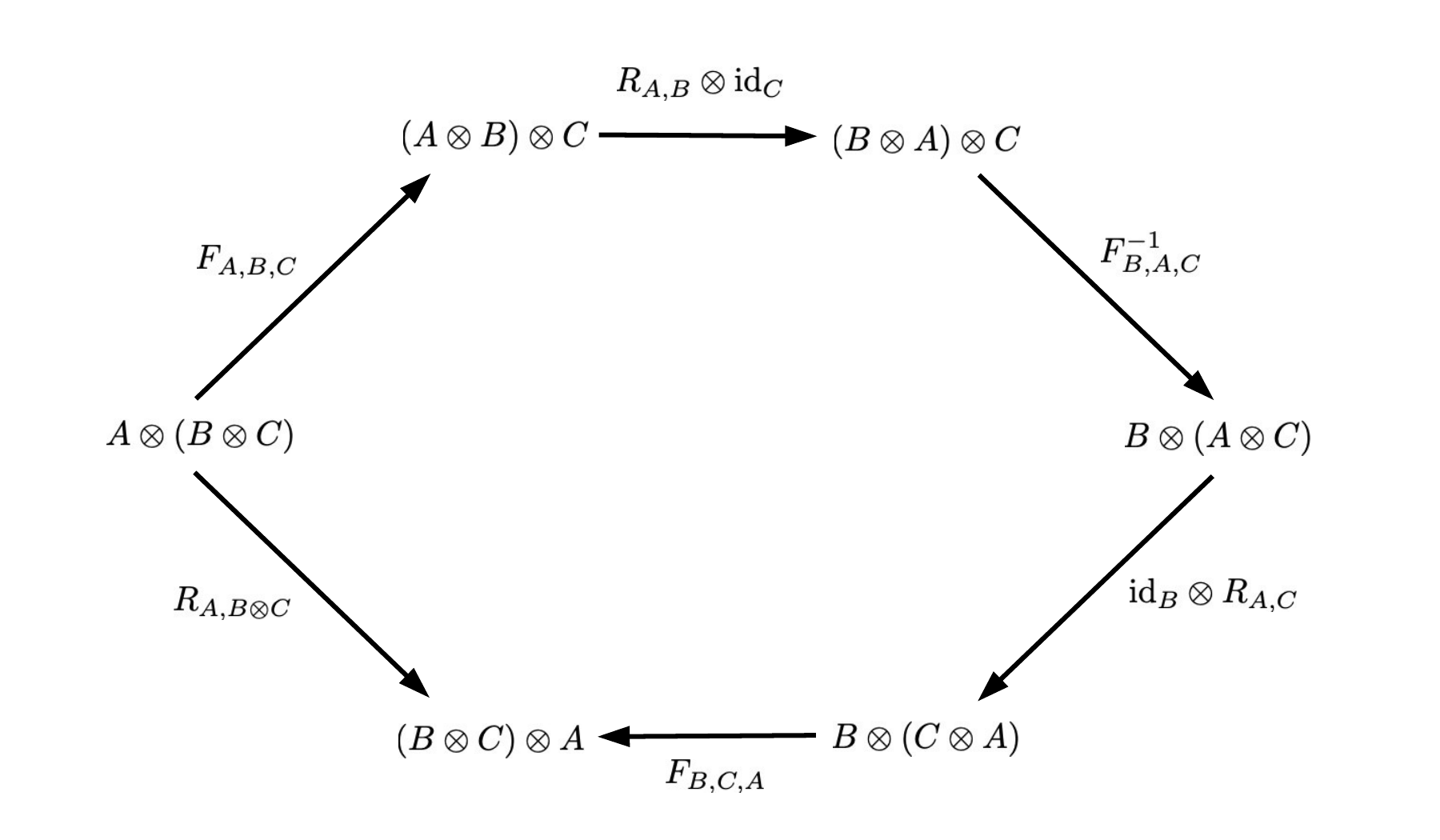}
  \caption{The hexagon identity.}
  \label{fig:Hexagon}
\end{figure}

A \textit{braiding} in a tensor category is a family of functorial isomorphisms\footnote{In mathematical literature it is usually denoted $c_{AB}$ and is called the commutativity constraint.},
\begin{equation}
    R_{A,B} : A \otimes B \rightarrow B \otimes A,
\end{equation}
such that the diagram in Figure \ref{fig:Hexagon}
and a similar diagram that starts with $(A \otimes B) \otimes C$ commute. A \textit{twist} of a braided tensor category is a family of natural isomorphisms:
\begin{equation}
    \theta_A : A \rightarrow X,
\end{equation}
compatible with the tensor product such that for each $A, B, C \in \CC$,
\begin{equation}
    \theta_{A \otimes B} = R_{B,A} \circ R_{A,B} (\theta_A \otimes \theta_B),
\end{equation}
\[ \begin{tikzcd}[row sep=large, column sep = large]
A \otimes B \arrow{r}{\theta_{A \otimes B}} \arrow[swap]{d}{\theta_A \otimes \theta_B} & A \otimes B\\
A \otimes B \arrow{r}{R} & B \otimes A \arrow{u}{R} 
\end{tikzcd},
\]
and with the duality operation,
\begin{equation}
    \theta_{A^*} = {\theta_A}^*.
\end{equation}
\paragraph{Example: Abelian MTC.}
In the case of Abelian MTCs, line operators (objects in the category) are parametrized by vectors $\vec{\alpha}$ in a lattice equipped with a quadratic form, an example of this is reviewed in Section \ref{sec:MTCM3}. In this case, the fusion is trivial\footnote{The true mathematical story behind trivial braiding is more complicated, but we ignore it here.} 
\begin{equation}
    (A \otimes B) \otimes C = A \otimes (B \otimes C).
\end{equation}
When a line $\vec{\alpha}$ braids around a line $\vec{\beta}$, the corresponding wavefunction picks up a $U(1)$ phase. This is the physical meaning of the braiding and amounts to the multiplication by
\begin{equation}
	R_{\vec{\alpha} \vec{\beta}} \equiv \exp\left[\pi i (h_{\vec{\alpha} + \vec{\beta}} - h_{\vec{\alpha}} - h_{\vec{\beta}})\right].
\end{equation}
Here, $h_{\alpha}$ is the conformal dimension of the corresponding line operator (or primary field $\Phi_{\alpha}$) or equivalently the \textit{topological spin} of the anyon $\vec{\alpha}$: 
\begin{equation}
	h_{\vec{\alpha}} \equiv \half \langle \alpha,\alpha\rangle,
\end{equation}
where the quadratic form $\langle \cdot , \cdot \rangle$ and the lattice are the data defining the category. We then see that
\begin{equation} \label{eq:topspin}
	\left(R_{\vec{\alpha},\vec{\beta}}\right)^2 = \frac{\theta(\vec{\alpha}+\vec{\beta})}{\theta(\vec{\alpha})\theta(\vec{\beta})} , \quad \theta(\vec{\alpha}) \equiv \exp\left[2\pi i h_{\vec{\alpha}} \right],
\end{equation}
where $\theta$ is known as the \textit{quadratic refinement} of the bilinear form $R$. We just described the modular tensor category (MTC) corresponding to an Abelian Chern-Simons theory.

\begin{figure}[h!]
\centering
	\includegraphics[width=0.8\textwidth]{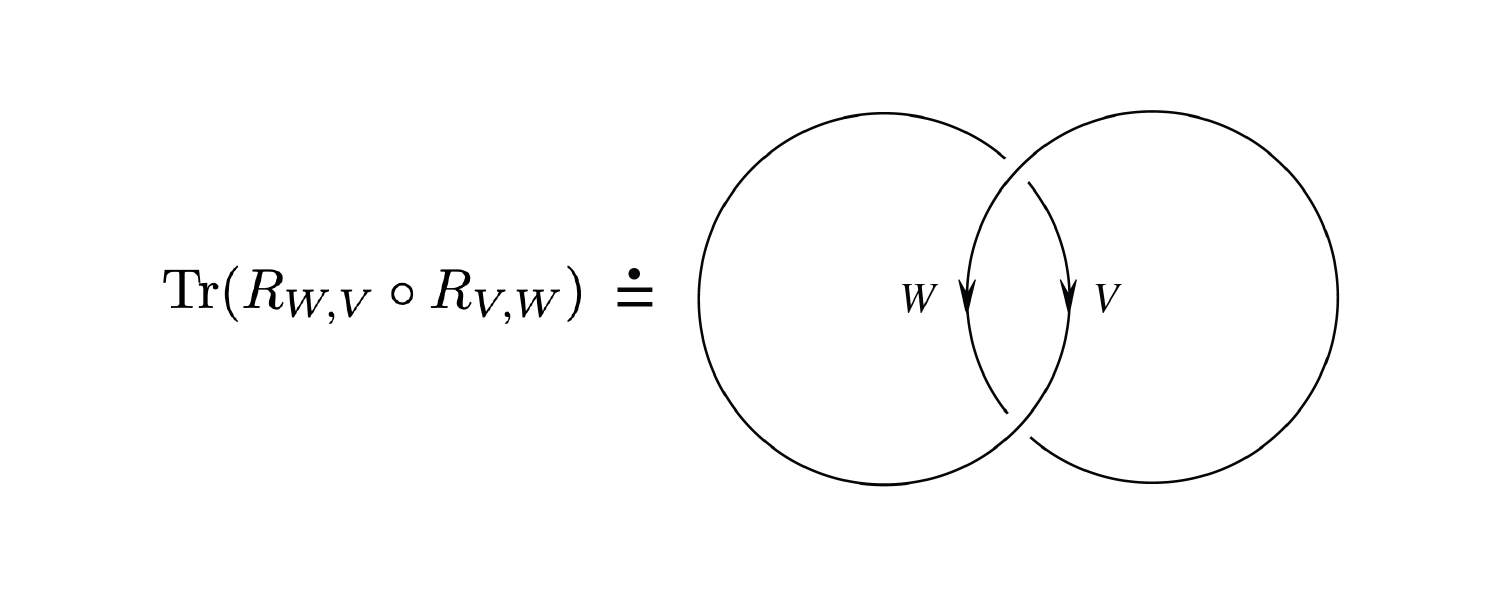}
  \caption{The S-matrix}
  \label{fig:Smatrix}
\end{figure}

\paragraph{Modular Tensor Category.}
A rigid tensor category with compatible braiding and twist is called a \textit{ribbon category}. For a morphism $f \in \mathrm{Hom}(A,A)=\mathrm{End}(A) $, define its \text{trace} in the following manner:
\begin{equation}
    \mathrm{Tr}(f) \equiv \mathrm{ev}_A \circ R_{A,A^*} \circ (\theta_A f \otimes \iden_{A^*})\circ {\mathrm{ev}_A}^*,
\end{equation}
which is in $\mathrm{End}(\iden)$ and hence a $k$-number. In particular,
\begin{equation}
    d_A \equiv \mathrm{dim}(A) = \mathrm{Tr}(\iden_A).
\end{equation}
In the case of the example of the Abelian MTC discussed above, $d_A = 1$ for all $A \in \mathcal{I}$.
A \textit{modular tensor category} is a semi-simple ribbon category with an invertible S-matrix:
\begin{equation}
    S_{i,j} = \mathrm{Tr}(R_{A_i,A_j} \circ R_{A_i,A_j}),
\end{equation}
where $A_i, A_j \in \mathcal{I}$, see Figure \ref{fig:Smatrix}. Here we choose representations of isomorphism classes of simple objects and enumerated them.
Define the diagonal T-matrix as follows\footnote{Note that $\theta_{A_i} \in \mathrm{End}(A_i)$ is a $k$-number for simple $A_i$.}
\begin{equation}
    T_{i,j} = \delta_{i,j} \theta_{A_i}.
\end{equation}
Then one can verify that the matrices $\{S,T\}$ form a projective representation of the modular group $SL(2,\IZ)$ \cite{bakalov2001lectures,RowellTensor}. 

\subsection{The Drinfeld associator and KZ-equations}
\label{sec:Drinfeld}

Let $\mathbb{C}[\mathcal{X},\mathcal{Y}][[\hbar]]$ be the ring freely generated by elements $\mathcal{X}, \mathcal{Y}$, complete over formal power series in $\hbar$. Consider the differential equation
\begin{equation} \label{eq:KZ}
    \Psi'(z) = \hbar\left(\frac{\mathcal{X}}{z} + \frac{\mathcal{Y}}{z-1}\right) \Psi(z),
\end{equation}
on the Riemann sphere, where solutions are understood as elements of $\mathbb{C}[\mathcal{X},\mathcal{Y}][[\hbar]]$ with coefficients depending on $z$. This differential equation has regular singularities at $z=0, 1, \infty$. Define fundamental solutions $\Psi_0$ and $\Psi_1$,
\begin{eqnarray}
    \Psi_0(z) & = & P(z) z^{\hbar \mathcal{X}}, \\
    \Psi_1(z) & = & Q(z) (1-z)^{\hbar \mathcal{Y}},
\end{eqnarray}
where $P(z)$ is analytic at $z=0$ with $P(0)=1$ and $Q(z)$ is analytic at $z=1$ with $Q(1)=1$. Define the transition function $\Phi(\mathcal{X},\mathcal{Y})$ between $\Psi_0$ and $\Psi_1$ as 
\begin{equation}
    \Psi_0(z) = \Psi_1(z) \Phi(\mathcal{X},\mathcal{Y}).
\end{equation}
It is not difficult to compute the first few coefficients of $\Phi$:
\begin{equation} \label{eq:asso}
    \Phi(\mathcal{X},\mathcal{Y}) = 1 - \frac{\zeta(2)}{(2\pi \sqrt{-1})^2}[\mathcal{X},\mathcal{Y}] \hbar^2 + \frac{\zeta(3)}{(2\pi \sqrt{-1})^3}\left([[\mathcal{X},\mathcal{Y}],\mathcal{Y}] - [\mathcal{X},[\mathcal{X},\mathcal{Y}]]\right)\hbar^3 + \mathcal{O}(\hbar^4),
\end{equation}
where $\zeta$ is the Riemann-zeta function. The importance of $\Phi(\mathcal{X},\mathcal{Y})$ was pointed out in \cite{Drinfeld:1989st} by Drinfeld where he used this function to define a tensor category structure on the category of finite dimensional $\frak{g}$-modules as follows. 

Let $\frak{g}$ be a finite dimensional simple Lie algebra and $(\cdot, \cdot)$ be its Killing form. The Killing form is an element of $\frak{g}^* \otimes \frak{g}^*$ and gives a linear isomorphism $\frak{g} \rightarrow \frak{g}^*$. Denote its image in $\frak{g} \otimes \frak{g}$ by $\Omega$. If we choose a linear, orthonormal with respect to the Killing form, basis $\{e_i\}$ in $\frak{g}$, we have
\begin{equation} \label{eq:OmegaAB}
    \Omega = \sum_{i=1}^{\dim \frak{g}} e_i \otimes e_i.
\end{equation}
For three finite dimensional $\frak{g}$-modules $A, B, C \in \gmod$ denote by $\Omega_{AB}$ the evaluation of $\Omega$ in $A \otimes B$, $\Omega_{BC}$ the evaluation of $\Omega$ in $B \otimes C$ and denote by the same letters the natural embedding of $\Omega_{AB}$, $\Omega_{BC}$ into $\mathrm{End}(A \otimes B \otimes C)$. Then \cite{Drinfeld:1989st} the elements 
\begin{equation} \label{eq:KZfusion}
    F_{A,B,C} = \Phi(\Omega_{AB},\Omega_{BC}) : A \otimes (B \otimes C) \rightarrow (A \otimes B) \otimes C,
\end{equation}
define a fusion structure on the category $\gmod[[\hbar]]$, i.e. they satisfy the pentagon axiom shown in Figure \ref{fig:Pentagon}. Equation \eqref{eq:KZ} with $\mathcal{X}=\Omega_{AB}$ and $\mathcal{Y}=\Omega_{BC}$ is known as the \textit{Knizhnik-Zamolodchikov equation} \cite{KNIZHNIK198483} (or in short KZ-equation) associated to the Lie algebra $\frak{g}$.

\paragraph{Monodromy and braiding.}
Because $\Psi_0(z) \sim z^{\hbar \mathcal{X}}$ as $z \rightarrow 0$, the monodromy of $\Psi_0$ around zero is $e^{2\pi i \hbar \mathcal{X}}$. Hence the half-monodromy around zero is
\begin{eqnarray}
    B_0 = e^{\pi i \hbar \mathcal{X}}.
\end{eqnarray}
Similarly, from $\Psi_1(z) \sim (1-z)^{\hbar \mathcal{Y}}$ when $z \rightarrow 1$, we obtain the half-monodromy along a small circle around $z=1$
\begin{equation}
    B_1 = \Phi(\mathcal{X},\mathcal{Y})^{-1} e^{\pi i \hbar \mathcal{Y}} \Phi(\mathcal{X},\mathcal{Y}).
\end{equation}
Thus, the braiding is defined by
\begin{equation} \label{eq:KZbraiding}
    R_{AB} = P_{AB} \circ \exp\left(i \pi \hbar \Omega_{AB}\right),
\end{equation}
where 
\begin{equation}
    P_{AB} : A \otimes B \rightarrow B \otimes A
\end{equation}
is the permutation of factors in the tensor product, $u \otimes v \mapsto v \otimes u$. It is easy to see \cite{Drinfeld:1989st} that \eqref{eq:KZbraiding} defines a braiding for the category of finite dimensional $\frak{g}$-modules over $\mathbb{C}[[\hbar]]$ with the fusion \eqref{eq:KZfusion}. 

\section{Rozansky-Witten theory}
\label{sec:RWtheory}

In this section we present the analysis of Rozansky-Witten theory as a braided tensor category. We start with the well-known Feynman diagram expansion of the theory and from there deduce the Lie algebra structure underlying the braiding morphism. We discuss conformal blocks and the associator morphism, invertible and non-invertible objects in the category, and finally discuss autoequivalences.

\subsection{Feynman diagram expansion}

\begin{figure}[h!]
\centering
	\includegraphics[width=0.6\textwidth]{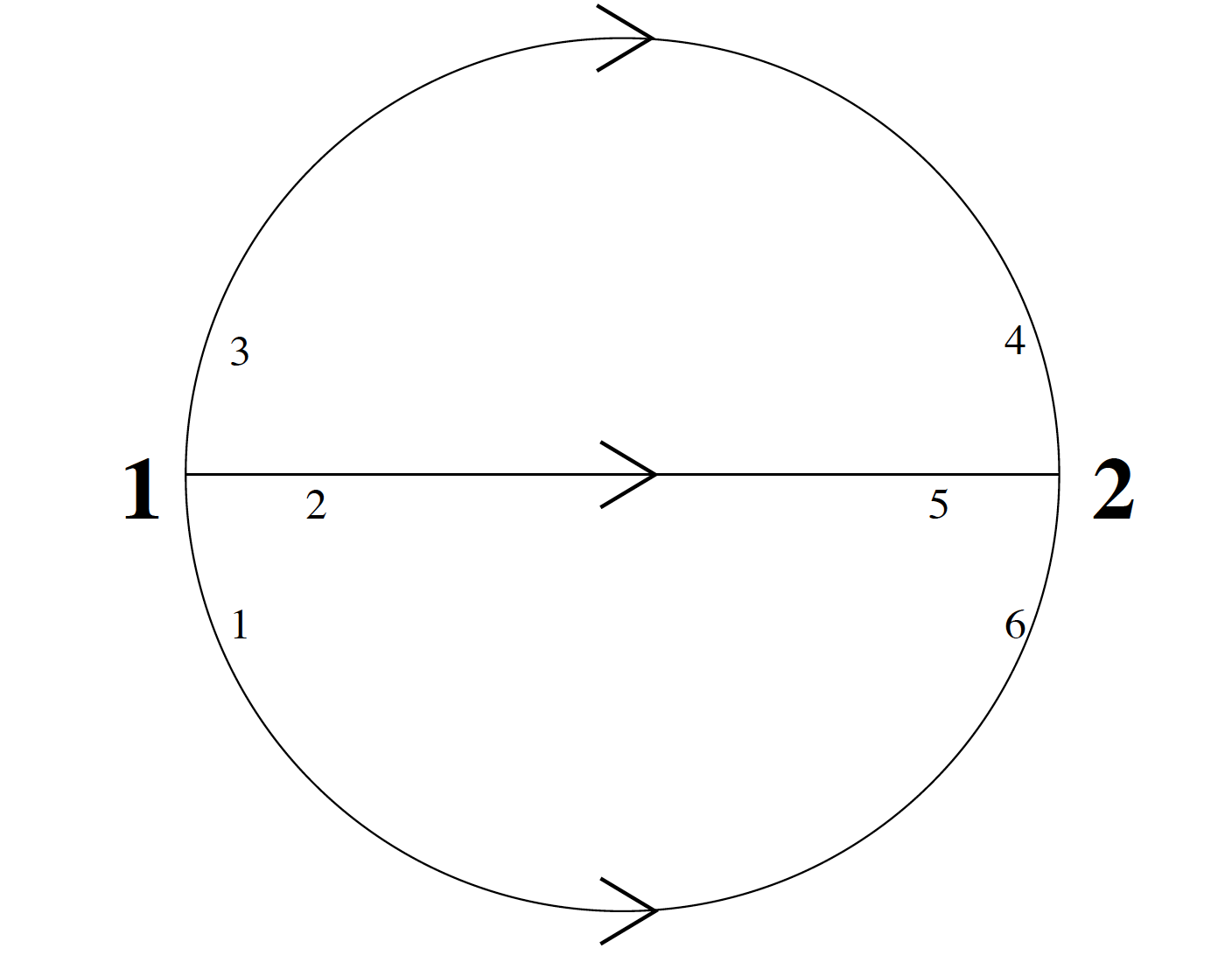}
  \caption{A typical diagram with two vertices.}
  \label{fig:diagram}
\end{figure}

 The partition function of RW-theory evaluated on a closed 3-manifold $M$ is a topological invariant. For compact target $X$, it can be evaluated via a Feynman diagram expansion which takes the following general form
\begin{equation}
    \sum_{\Gamma} b_{\Gamma}(X) I^{\mathrm{RW}}_{\Gamma}(M),
\end{equation}
where $\Gamma$ denotes trivalent graphs. The quantities $b_{\Gamma}(X)$ are known as \textit{weights} and solely depend on the hyperk\"ahler manifold $X$. We note here that for a manifold $X$ of dimension $4k$ the weights $b_{\Gamma}(X)$ vanish except when $\Gamma$ has $2k$ vertices. A typical diagram has the following form depicted in Figure \ref{fig:diagram}.
On the other hand the number $I^{\mathrm{RW}}_{\Gamma}(M)$ depends solely on the 3-manifold $M$. 

Now, consider a closed $3$-manifold $M$ with a link $\mathcal{L}$ in it. First, let us review the situation for Feynman diagram expansions of ordinary Chern-Simons theory with Wilson lines. The perturbative expansion around the trivial gauge connection gives rise to 3-manifold invariants of the form
\begin{equation}
    \sum_{\Gamma} c_{\Gamma}(\mathfrak{g};V_a) Z_{\Gamma}(M;\mathcal{L}),
\end{equation}
where the weights $c_{\Gamma}$ now depend on the Lie algebra $\mathfrak{g}$ of the Chern-Simons gauge group and the Wilson line representations $V_a$. The invariants $Z_{\Gamma}(M;\mathcal{L})$ solely depends on the link $\mathcal{L}$ embedded in the 3-manifold $M$. The sum is taken over all Feynman diagrams with external vertices on Wilson lines and with fixed Euler characteristic $\chi(\Gamma)$. For a link with $m$ components the space of diagrams is denoted by $\mathbf{\mathcal{A}}(S^1 \cup \ldots \cup S^1)$, where one takes the disjoint union of $m$ copies of $S^1$. It is then believed that the terms  $Z_{\Gamma}(M;\mathcal{L})$ are given by the coefficients of the framed Kontsevich integral
\begin{equation}
    Z^{\mathrm{Kont}}(M;\mathcal{L}) = \sum_{\Gamma} Z_{\Gamma}^{\mathrm{Kont}}(M;\mathcal{L}), \quad \Gamma \in \mathbf{\mathcal{A}}(S^1 \cup \ldots \cup S^1)
\end{equation}
of the link $\mathcal{L}$ in $M$. In the following, we will be interested in the weight system $c_{\Gamma}(\mathfrak{g};V_a)$. The data consists of the Lie algebra structure on $\frak{g}$, an invariant scalar product
\begin{equation}
    (\cdot, \cdot) : \frak{g}^{\otimes 2} \rightarrow k,
\end{equation}
and $\frak{g}$ and representations $\{V_a\}$ of $\frak{g}$ coloring connected components of $\mathcal{L}$. Recall that a representation is a pair $(V_a, \rho_a)$ where 
\begin{equation}
    \rho_a : \frak{g} \rightarrow \mathrm{End}(V_a)
\end{equation}
is a Lie algebra homomorphism. If we choose a basis $\{x_i\}$ in $\frak{g}$, the Lie albebra structure is represented by structural constants 
\begin{equation}
    [x_i, x_j] = c^k_{~ij} ~x_k,
\end{equation}
the scalar product by the matrix
\begin{equation}
    \sigma_{ij} = (x_i, x_j),
\end{equation}
and if we choose a basis $\{e_{K_a}\}_{K_a = 1}^{\dim(V_a)}$ in $V_a$, then $\rho_a$ is represented by the matrix $\rho_a(x_i)_{L_a}^{K_a}$. For the $\theta$ graph from Figure \ref{fig:diagram}, where all edges are colored by the adjoint representation, we have 
\begin{equation}
    c_{\Gamma}(\frak{g},R) = \sum_{i,j,k} c^k_{~ij} \sigma_{kl} \sigma^{im} \sigma^{js} c^l_{~ms}.
\end{equation}
Here $\sigma^{ij}$ is the inverse matrix to $\sigma_{ij}$. 
For the chord diagram of the half-circle one then gets for example
\begin{equation}
    c_{\Gamma}(\mathfrak{g};R) = \sum_{i,j,K,L} \rho_{iK}^{~~L} \rho_{jL}^{~~K} \sigma^{ij}.
\end{equation}

\begin{figure}[h!]
\centering
\includegraphics[width=0.3\textwidth]{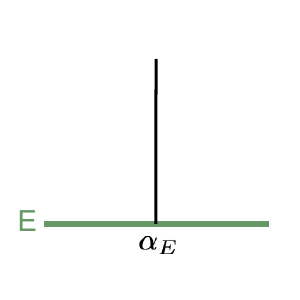}
  \caption{Diagramatic representation of the Atiyah Class.}
  \label{fig:Atiyah}
\end{figure}

The situation for RW-theory is now very analogous to the above. The dictionary is then as follows
\begin{align}
    c_T \in \Omega^{0,1}(X,\mathrm{Sym}^3 T^*) \quad & \leftrightarrow \quad c_{ijk} \nonumber \\
    \tilde \omega \in H^0(X, \Lambda^2 T) \quad & \leftrightarrow \quad \sigma^{ij} \nonumber
\end{align}
where $c_T$ and $\tilde \omega$ are the Riemannian curvature and the dual of the holomorphic symplectic form $\omega$ respectively. In the following, we will choose holomorphic vector bundles on $X$ labeled by $E_a$. Further, we require that each such $E$ is equipped with a smooth hermitian metric $h$ along fibers and with a connection $\nabla$ that is 
compatible with $h$. This means that for all sections $s, t \in \Omega^0(X;E)$, the following identity holds:
\begin{equation}
    dh(s,t) = h(\nabla s, t) + h(s, \nabla t).
\end{equation}
In the RW theory, Wilson line observables in the \textit{adjoint} representation are given by
\begin{equation}
    W_C(T^{1,0} X) = \textrm{Tr} \mathbf{P} \exp \int_C \mathcal{A}, \quad \textrm{where } \mathcal{A}^i_j = - d\phi^k \Gamma^i_{kj} + \eta^{\bar i} \chi^k {(\alpha_T)^i_j}_{\bar i k},
\end{equation}
where $\alpha_T$ is the curvature of the connection that comes with the holomorphic tangent bundle. Wilson line observables in a representation corresponding to holomorphic vector bundles $E \rightarrow X$ on $X$ are
\begin{equation}
    W_C(E) = \mathrm{Tr} \mathbf{P} \exp \int_C \mathcal{A}, \quad \textrm{where } \mathcal{A} = - d\phi^k A_k + \eta^{\bar i} \chi^k (\alpha_E)_{\bar i k},
\end{equation}
where $\alpha_E$ denotes the curvature of the connection on $E$ and is further specified below. In order to define RW analogs of representations $(V_a,\rho_a)$, one proceeds as follows. For an $m$-component link, one chooses holomorphic vector bundles $E_1,\ldots, E_m$ over $X$, of rank $r_1, \ldots, r_m$, respectively. Let the curvatures of the corresponding connections $\nabla_a$ be
\begin{equation}
    \alpha_{E_a} \in \Omega^{1,1}(\mathrm{End}E_a),
\end{equation}
or in local complex coordinates they are given by the matrices
\begin{equation}
    (\alpha_a)^{I_a}_{J_a k\bar l} d z_k \wedge d \bar{z}_l.
\end{equation}
Then we have the correspondence 
\begin{equation}
    \alpha_{E_a} \quad \leftrightarrow \quad (V_a,\rho_a).
\end{equation}
$\alpha_E$ is a representative of the \textit{Atiyah class} $[\alpha_E]$ of the bundle $E$ and can be viewed as a map
\begin{equation}
    \alpha_E: E \otimes T \rightarrow E[1],
\end{equation}
when $E$ is an object in the derived category, and is pictorially represented as in Figure \ref{fig:Atiyah}.
We further introduce $\alpha_T \in \Omega^{0,1}(T^* \otimes \mathrm{End}(T))$ as the curvature of the connection on the tangent bundle. Then for vector fields $t_1, t_2, t_3 \in \Gamma(T)$\footnote{Here, $\Gamma(E)$ denotes the space of local sections of a holomorphic vector bundle $E$.}, define
\begin{equation}
    c_T(t_1,t_2,t_3) = \omega(\alpha_T(t_1,t_2),t_3).
\end{equation}
Here, $\omega$ is the holomorphic symplectic form on $X$. Moreover, there is an identity between $\alpha_E$ and $\alpha_T$ which for vector fields $t_1, t_2 \in \Gamma(T)$ and $e \in \Gamma(E)$ reads as follows,
\begin{equation}
    \alpha_E(t_1,\alpha_E(t_2,e)) - \alpha_E(t_2,\alpha_E(t_1,e)) = \alpha_E(\alpha_T(t_1,t_2),e) + \bar{\partial}(\ldots). \label{eq:STU}
\end{equation}
This identity holds up to a $\bar{\partial}$-derivative and will become exact once we pass to cohomology classes. It is known as the \textit{STU-relation} and has the following graphical representation.
\begin{figure}[h!]
\centering
\includegraphics[width=\textwidth]{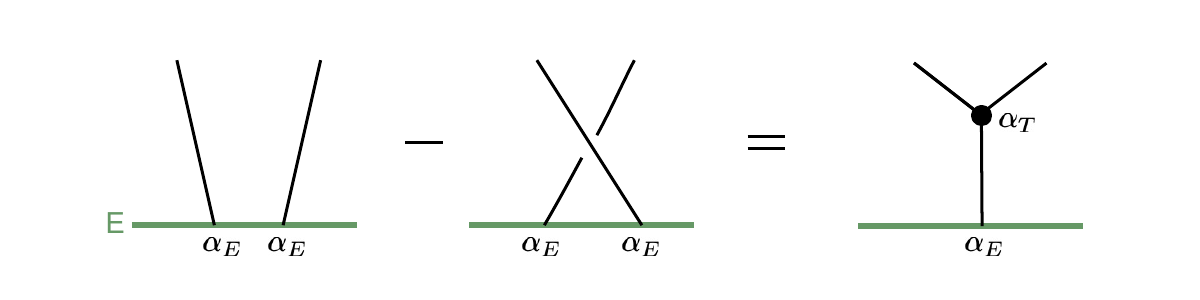}
  \caption{Diagramatic representation of the STU relation.}
  \label{fig:STU}
\end{figure}

\subsection{Space of boundary states}
\label{sec:hilbspace}

Let $\Sigma_{g,m}$ be a compact oriented surface of genus $g$ with $m$ punctures labeled by holomorphic vector bundles $E_1, \ldots, E_m$ over $X$. Here we will describe the space of boundary states for RW topological quantum field theory and will focus on the case $g=0$. There is a one-to-one correspondence between the states of boundary states and the space of ``conformal blocks'' as will be explained in section \ref{sec:RWbraiding}, and hence we sometimes refer to these spaces as conformal blocks in the following.

For $m=0$, the Hilbert space is
\begin{equation}
\mathcal{H}_{RW[X]} (S^2) = H^{0, \bullet} (X) = \oplus_n H^{n} (\mathcal{O}_X)
\end{equation}
The corresponding Verlinde formula that gives the dimension of this space is obtained by taking the (super-)trace, {\it i.e.} evaluating the invariant on $M_3 = S^1 \times S^2$:
\begin{equation}
\text{sdim} \mathcal{H}_{RW[X]} (S^2) = \sum_i (-1)^i \dim H^{0, i} (X)
\end{equation}
A generalization of this Verlinde formula to $m \ne 0$ looks like, {\it cf.} \cite{Qiu:2020mji}:
\begin{equation}
\text{sdim} \mathcal{H}_{RW[X]} (S^2\backslash\mathrm{pts.}; E_1,\ldots, E_m) = \chi (E_1 \otimes \ldots \otimes E_m)
\end{equation}
and the corresponding space of boundary states are
\begin{equation} \label{eq:HilbS2}
\mathcal{H}_{RW[X]} (S^2\backslash\mathrm{pts.}; E_1,\ldots, E_m)
= \bigoplus_{n=0}^{\dim_{\mathbb{C}} X}
H^n_{\bar \partial} (X, E_1 \otimes \ldots \otimes E_m)
\end{equation}
This data is basically the structure of morphisms in the category underlying the Rozansky-Witten TQFT (see {\it e.g.} \cite{MR2661534,Gukov:2020lqm}). This is a monoidal category, with the unit object given by the structure sheaf $\mathcal{O}_X$; it corresponds to the ``transparent'' line operator. And the braiding structure (or, a suitable variant of it) is precisely what we are trying to explore here. The analogue of the Lie algebra is the space of vector fields, with the product given by the Atiyah class as explained in the following.

\subsection{Lie algebra structure}

We find that for RW-theory there are two types of Lie algebras one can define. One goes back to Kapranov \cite{Kapranov}, while the other has its roots in \cite{MR2661534}. Both two Lie algebra instantiations have virtues and disadvantages, and while one is more intuitive when dealing with cohomology groups, the other is more intuitive in sheaf theory. But ultimately, what enters the physics is not the concrete form of the Lie algebra but rather the Lie bracket and the Cartan Killing form. Let us, for the sake of completeness, give an overview of these two Lie algebras. In the sheaf theoretic case, one considers the holomorphic tangent bundle $T$ as a sheaf over $X$\footnote{Note that this sheaf, while having local sections, might not have global sections over $X$.}. Viewed as an object in the category of coherent sheaves, we fix an open covering $\cup_{\alpha} U_{\alpha}$ of $X$ such that it trivializes $T$ as follows:
\begin{equation}
    \phi_{\alpha} : \mathcal{T}(U_{\alpha}) \rightarrow \mathcal{O}_X(U_{\alpha})^{\oplus n}.
\end{equation}
Thus over each patch holomorphic sections admit a finite basis as $\mathcal{O}_X$ modules. Choosing then over each patch a particular section, then gives rise to a tuple of local sections
\begin{equation}
    (s_1, s_2, \ldots, s_{\alpha},\ldots).
\end{equation}
Such a tuple can be viewed as an element of our Lie algebra while to be fully precise we should take sections of the shifted tangent bundle $T[-1]$. Then the Lie bracket is given by the Atiyah class $\alpha_T$ viewed as a map given in equation \eqref{eq:L1Atiyah}. The Cartan Killing form is then given in terms of the homorphic 2-form $\omega$ which for Calabi-Yau manifolds, e.g. $X=K3$, admits a unique global section. Note that $\alpha_T$ defines a Lie algebra structure on $\Gamma(T)$ up to $\bar{\partial}$. Similarly, $\alpha_E$ defines on $\Gamma(E)$ the module over this Lie algebra, also up to $\bar{\partial}$. The Jacobi identity is then the STU relation \eqref{eq:STU}. The Lie algebra $\Gamma(T)$ and the module $\Gamma(E)$ are finite dimensional over $\mathcal{O}_X$. 

The second viewpoint arises when passing to the spaces of Dolbaux cohomologies. Here, the above structures become Lie algebras and modules over Lie algebras. Concretely, one can view the Atiyah classes $\alpha_T$ and $\alpha_E$ as elements of $H^1(X,\textrm{Hom}(S^2 T,T))$ and $H^1(X,\textrm{Hom}(T \otimes E,E))$ where $S^2 T$ denotes the symmetrization of the tensor product of two tangent bundles. Then, following \cite{Kapranov}, for $A$ any quasicoherent sheaf of commutative $\mathcal{O}_X$-algebras, the maps
\begin{equation}
    H^i(X,T \otimes A) \otimes H^j(X, T \otimes A) \rightarrow H^{i+j+1}(X,T \otimes A), 
\end{equation}
given by composing the cup-product with $\alpha_T$ give the space $H^{\star -1}(X,T \otimes A)$ the structure of a graded Lie algebra\footnote{This structure is trivial, that is the bracket vanishes if we take $A=\mathcal{O}_X$ as pointed out in \cite{Kapranov}. To get a nontrivial bracket it is convenient to consider the algebra $S^m T^*$, that is the symmetric tensor powers of the cotangent sheaf.}. Similarly, for any holomorphic vector bundle $E$ on $X$ the maps
\begin{eqnarray}
    H^i(X,T \otimes A) \otimes H^j(X,E \otimes A) \rightarrow H^{i+j+1}(X,E \otimes A) \label{eq:Tmodule}
\end{eqnarray}
given by composing the cup-product with the class $\alpha_E$, give $H^{\star - 1}(X,E \otimes A)$ the structure of a graded $H^{\star-1}(X,T \otimes A)$-module. 

One can convince oneself that in both Lie algebras, the analogue of the operator $\Omega_{AB}$ defined in \eqref{eq:OmegaAB}, is then the operator 
\begin{eqnarray}
    H_{A,B} \in \mathrm{Ext}^2(A\otimes B, A\otimes B),
\end{eqnarray}
which choosing local coordinates, can be written as 
\begin{equation}
    H_{A,B} = (\alpha_A \otimes \alpha_B)(\tilde{\omega}) = \sum_{k,m} \left(\alpha_A\right)^{I_A}_{~J_A k \bar l} \left(\alpha_B\right)^{I_B}_{~J_B m \bar n} \tilde{\omega}^{k m}, \label{eq:Hdef} 
\end{equation}
The module structure defined in \eqref{eq:Tmodule} then gives an action of $H_{A,B}$ on the state space \eqref{eq:HilbS2}. This action can be nontrivial as shown by the following example. \paragraph{Example.} Consider a sheaf $\mathcal{F}$ given by 
\begin{equation}
    \mathcal{F} = \mathcal{E}_1 \oplus \mathcal{E}_2.
\end{equation}
We will make no further assumptions on the subsheafs $\mathcal{E}_1$ and $\mathcal{E}_2$ except that there should exist a homomorphism
\begin{equation}
    f : \mathcal{E}_1 \rightarrow \mathcal{E}_2.
\end{equation}
An example is to take the structure sheaf $\mathcal{O}_X$ and the sheaf $\mathcal{O}_C$ for a holomorphic curve $C$ on $X$, with the map $f: \mathcal{O}_X \rightarrow \mathcal{O}_C$ being the restriction map. Then the dual to $f$ is a degree two map
\begin{equation} \label{eq:fvee}
    f^{\vee} : \mathcal{E}_2 \rightarrow \mathcal{E}_1[2]. 
\end{equation}
We then consider the space of boundary states on the two-punctured sphere given by
\begin{equation}
    \mathcal{H}_{RW[X]}(S^2;\mathcal{F},\mathcal{F}) = \bigoplus_{n=0}^{\mathrm{dim}_{\mathbb{C}}X} H_{\bar{\partial}}^n(X,\mathcal{F} \otimes \mathcal{F}),
\end{equation}
which can be computed by noting that
\begin{equation}
    \mathcal{F} \otimes \mathcal{F} = \mathcal{E}_1^{\otimes 2} \oplus (\mathcal{E}_1 \otimes \mathcal{E}_2)^{\oplus 2} \oplus \mathcal{E}_2^{\otimes 2},
\end{equation}
and hence\footnote{We thank Will Donovan for pointing out this example to us.}
\begin{equation} \label{eq:HilbFF}
    \mathcal{H}_{RW[X]}(S^2;\mathcal{F},\mathcal{F}) = H^*(\mathcal{E}_1^{\otimes 2}) \oplus H^*(\mathcal{E}_1\otimes \mathcal{E}_2)^{\oplus 2} \oplus H^*(\mathcal{E}_2^{\otimes 2}).
\end{equation}
Then $H_{\mathcal{F},\mathcal{F}}$ acts nontrivially on the state space $\mathcal{H}_{RW[X]}$ given in \eqref{eq:HilbFF}. This action has, apart from a diagonal component (when considered as an action on the components in the sum) which will be explored later on, a non-diagonal component  given by $f^\vee$ in \eqref{eq:fvee}.

\subsection{Braiding}
\label{sec:RWbraiding}

\begin{figure}[h!]
\centering
\includegraphics[width=0.3\textwidth]{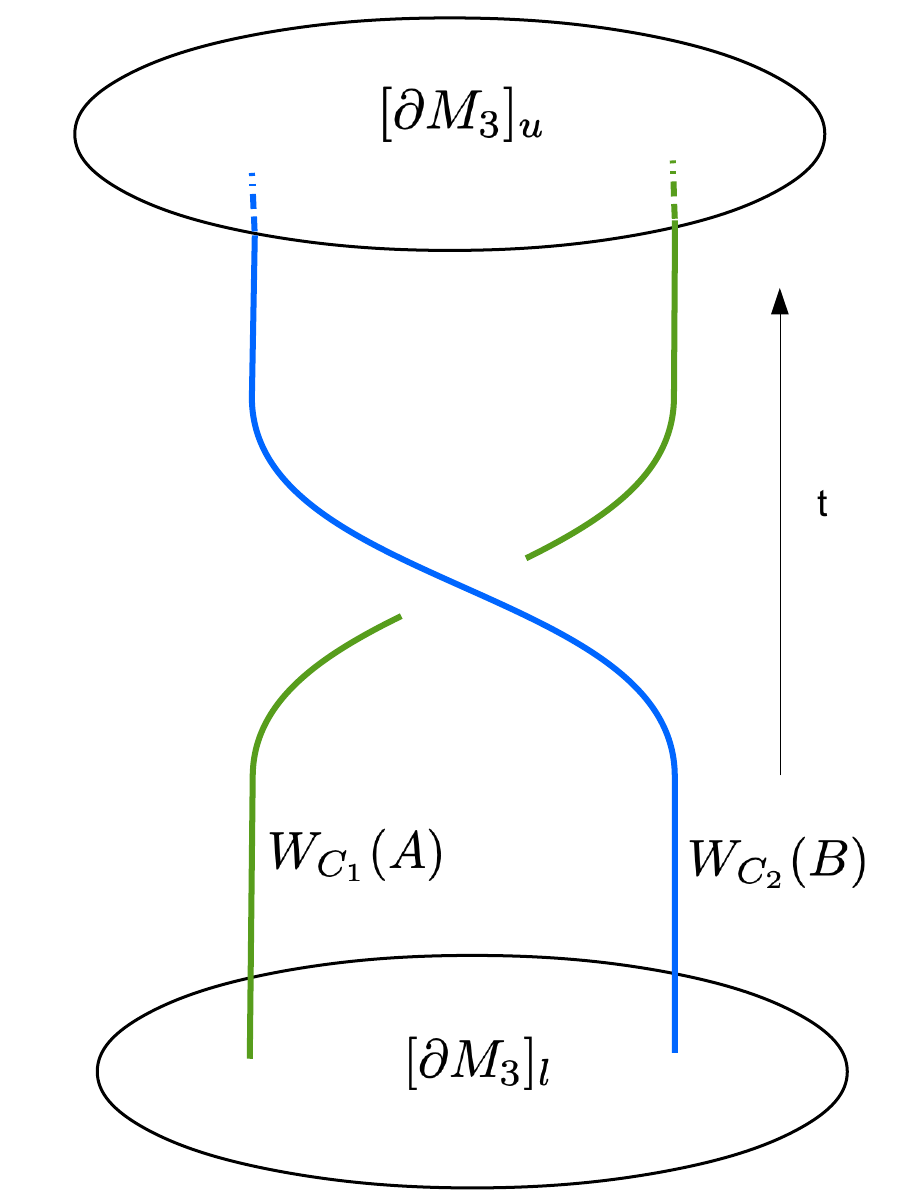}
  \caption{Insertion of two braided Wilson lines puncturing the boundary of a 3d cylinder at its two ends.}
  \label{fig:Bamp}
\end{figure}

We have seen that, as was explained earlier, line observables correspond to sheaves on the manifold $X$. These observables are parametrized by objects of the derived category $D^b(X)$ of coherent sheaves of $X$. In order to compute braiding, we can follow two different approaches. The first one, which is more natural from a physics viewpoint, is to compute the transition amplitude
\begin{equation}
    \prescript{}{\mathcal{H}_u}{\langle} B \otimes A | W_{C_1}(A) W_{C_2}(B) | A \otimes B \rangle_{\mathcal{H}_l},
\end{equation}
between two states in the Hilbert space corresponding to $\mathcal{H}([\partial M_3]_l, A \otimes B)$ and $\mathcal{H}([\partial M_3]_u, B \otimes A)$. Here $\mathcal{H}_u$ and $\mathcal{H}_l$ correspond to the Hilbert spaces attached to the upper/lower boundaries of a cylinder $M_3 = I \times \Sigma$. To do this, one proceeds by connecting the two punctures by line defects along the time direction as depicted in the Figure \ref{fig:Bamp}.
The corresponding transition amplitude can be computed using RW Feynman diagrammatic expansion of the path integral with appropriate boundary conditions.

\begin{figure}
\centering
\includegraphics[width=0.4\textwidth]{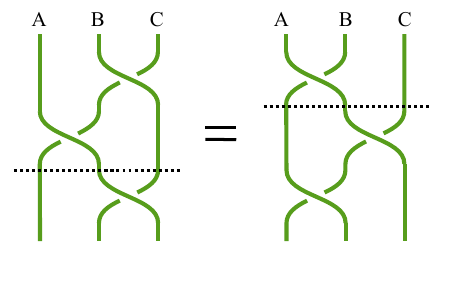}
  \caption{Reidemeister 3 move between three objects $A$, $B$ and $C$.}
  \label{fig:Reid3}
\end{figure}

Another approach, which we choose to follow here, is to define the braiding morphism as in \eqref{eq:KZbraiding} and the fusion using the Drinfeld associator \eqref{eq:KZfusion}. To be precise, given two objects, $A$, $B$ of $D(X)$, let $H_{A,B} \in \mathrm{Hom}_{D(X)}(A \otimes B, A \otimes B)$ and $C_A \in \mathrm{Hom}_{D(X)}(A,A)$ be two morphisms corresponding to the diagrams depicted below in Figure \ref{fig:Rtwist}.
\begin{figure}[h!]
\centering
\includegraphics[width=0.6\textwidth]{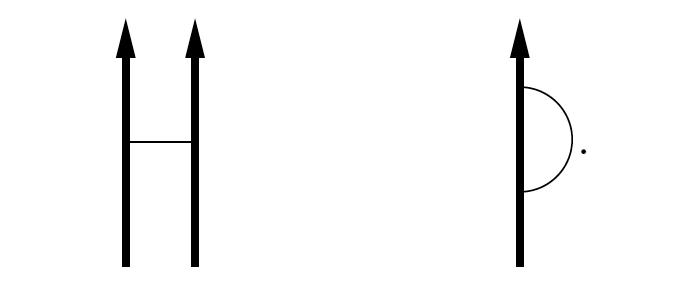}
  \caption{Diagrams corresponding to R-matrix and twist $\theta$.}
  \label{fig:Rtwist}
\end{figure}
Each is really an element of $\mathrm{Ext}^2$ and while $H_{A,B}$ was defined in \eqref{eq:Hdef}, $C_A$ is given by 
\begin{equation}
    C_A = (\alpha_A)^2 (\tilde{\omega}) = \sum_{k,m} \left(\alpha_A\right)^{I_A}_{~J'_A k \bar l} \left(\alpha_A\right)^{J'_A}_{~J_A m \bar n} \tilde{\omega}^{k m},
\end{equation}
that is $H_{A,B} \in \mathrm{Ext}^2(A\otimes B, A \otimes B)$ abd $C_A \in \mathrm{Ext}^2(A,A)$. $H_{A,B}$ is the RW analogue of $\Omega_{A,B}$ in section \ref{sec:Drinfeld}. In terms of these, the braiding morphism $R_{A,B}$ may be described as 
\begin{equation} \label{eq:braiding}
    R_{A,B} \equiv P_{A,B} \circ \exp\left( i\pi \hbar H_{A,B}\right)  \in \mathrm{Ext}^*(A\otimes B, B \otimes A) = \mathrm{Hom}_{D(X)}(A \otimes B, B \otimes A),
\end{equation}
where $P_{A,B}$ denotes the interchange of $A$ and $B$ in the tensor product. The twist is then defined as
\begin{equation} \label{eq:twist}
    \theta_A = \exp\left(i \pi \hbar C_A\right).
\end{equation}
\noindent
The expressions for the braiding and the twist can be viewed as Feynman diagram expansions\footnote{One might wonder why the Feynman diagram expansion does not contain trivalent vertices. The reason is that these can be eliminated using the STU relations for the tangent bundle and hence the only remaining diagrams are chord diagrams.} but here we would like to take them as definitions. As a 3d TQFT, correlation functions in the RW-theory should satisfy the Reidemeister III move depicted in Figure \ref{fig:Reid3} which in equations translates to the quantum Yang-Baxter equation:
\begin{eqnarray}
    ~ & ~ & F_{C,B,A}^{-1} \circ (R_{B,C} \otimes \mathrm{id}_A) \circ F_{B,C,A} \circ (\mathrm{id}_B \otimes R_{A,C}) \circ F_{B,A,C}^{-1} \circ (R_{A,B} \otimes \mathrm{id}_C) \nonumber \\
    ~ & = & (\mathrm{id}_C \otimes R_{A,B}) \circ F_{C,A,B}^{-1} \circ (R_{A,C} \otimes \mathrm{id}_B) \circ F_{A,C,B} \circ (\mathrm{id}_A \otimes R_{B,C}) \circ F_{A,B,C}^{-1}~. \label{eq:YB}
\end{eqnarray}

\paragraph{The associator.}
The associator $\Phi(\mathcal{X},\mathcal{Y})$ is written as a polynomial in the non-commuting variables
\begin{equation}
    \mathcal{X} \equiv H_{A,B} \otimes \mathrm{id}_C, \quad \mathcal{Y} \equiv \mathrm{id}_A \otimes H_{B,C}.
\end{equation}
The solution $\Psi$ of the KZ-equation \eqref{eq:KZ} is the analog of conformal blocks in Chern-Simons theory. Thus it can be viewed as a state in the Hilbert space and by choosing boundary conditions at $z=0$ these solutions can be identified with the states described in section \eqref{sec:hilbspace}. We can also argue that $\Psi$ is a horizontal section of the conformal block bundle with connection specified by \eqref{eq:KZ}. Self-consistency requires this connection to be flat which is a consequence of the identities
\begin{align}
    &\left [H_{A,B} \otimes \mathrm{id}_C + \mathrm{id}_A \otimes H_{B,C}, H_{A,C} \otimes \mathrm{id}_B \right ]=0 \label{eq:ACflat}\\
    &\left [H_{A,C} \otimes \mathrm{id}_B + \mathrm{id}_A \otimes H_{C,B},H_{A,B} \otimes \mathrm{id}_C  \right ]=0, \label{eq:ABflat}
\end{align}
which themselves are again a consequence of the STU relation (see Figure \ref{fig:STU}).
\paragraph{Example: K3 surface.}
Note that for $X$ a $K3$ surface, the associator becomes trivial as looking at formula \eqref{eq:asso}, we see that $\mathcal{X}$ and $\mathcal{Y}$ are both elements of $\mathrm{Ext}^2(E, E)$ with $E = A \otimes B \otimes C$, the commutator $[\mathcal{X},\mathcal{Y}]$ is an element of $\mathrm{Ext}^4(E,E)$ and hence vanishes for hyperk\"ahler manifolds of complex dimensions $2$. This shows that the associator is trivial for $X=K3$. Similarly, the quantum Yang-Baxter identity \eqref{eq:YB} becomes trivial as the expansion of the braiding operators only contributes up to order $\hbar$. To this order, left- and right-hand sides of \eqref{eq:YB} both equal $\hbar (H_{AB} +H_{BC} + H_{AC})$.

\begin{figure}[h]
\centering
\includegraphics[width=0.4\textwidth]{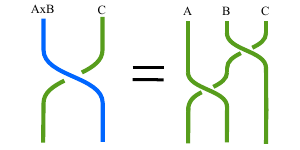}
  \caption{Braiding $A \otimes B$ with $C$ is equivalent to braiding first $A$ and then $B$ with $C$.}
  \label{fig:FusedBraiding}
\end{figure}

\begin{figure}[h]
\centering
\includegraphics[width=0.4\textwidth]{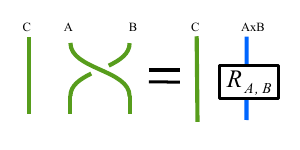}
  \caption{Braiding of $A$ with $B$.}
  \label{fig:FusedBraiding-b}
\end{figure}

\begin{figure}[h]
\centering
\includegraphics[width=0.5\textwidth]{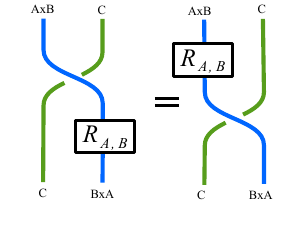}
  \caption{Equivalent version of the Reidemeister 3 move.}
  \label{fig:FusedReid3}
\end{figure}

\begin{figure}[h]
\centering
\includegraphics[width=0.6\textwidth]{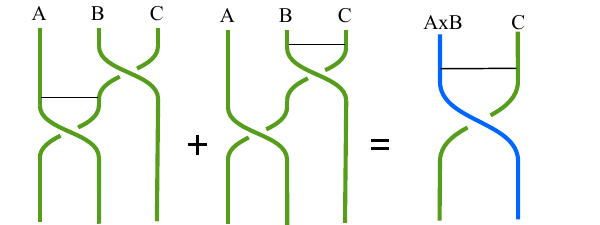}
  \caption{Braiding equivalence at first order in $\hbar$.}
  \label{fig:DoubleBraiding}
\end{figure}

\begin{figure}[h]
\centering
\includegraphics[width=0.6\textwidth]{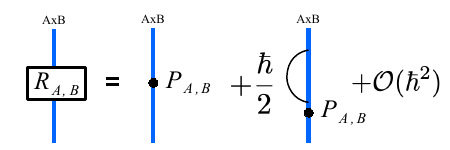}
  \caption{Expansion of the fused braiding morphism as a power series in $\hbar$.}
  \label{fig:FusedBraiding-c}
\end{figure}

\begin{figure}[h]
\centering
\includegraphics[width=0.6\textwidth]{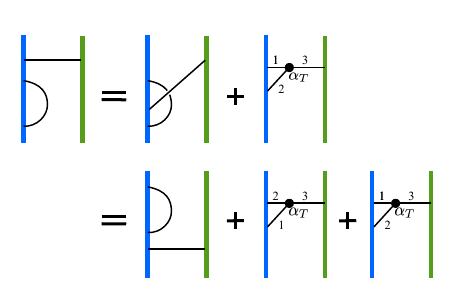}
  \caption{Commutation relation between $R_{A\otimes B,C}$ and $R_{A,B}$.}
  \label{fig:ProofYB}
\end{figure}

Next, we want to examine the Yang-Baxter identity \eqref{eq:YB} in other situations where the associators become trivial or do not contribute. In order to organize the computation better, as a first step, we will fuse the two lines $A$ and $B$ together and then succesively perform the braiding operations. The result is shown in Figure \ref{fig:FusedBraiding},
and Figure \ref{fig:FusedBraiding-b}, 
which represent the upper and lower part of Figure \ref{fig:Reid3}, respectively. Thus, in our diagramatic notation, the Reidemeister move boils down to sliding the $R_{A,B}$ morphism upwards and past the braiding between $A\otimes B$ with $C$, as shown in Figure \ref{fig:FusedReid3}.
Let us now realize the above identities explicitly in Rozanksky-Witten theory by using the expression of the braiding morphism given in equation \eqref{eq:braiding}. To begin with, we have the following identity of Atiyah classes,
\begin{equation}
    \iden_A \otimes \alpha_B + \alpha_A \otimes \iden_B = \alpha_{A \otimes B}.
\end{equation}
An immediate corollary of this is the diagrammatic identity at first order in $\hbar$ depicted in Figure \ref{fig:DoubleBraiding}.
Moreover, for the fused product $A\otimes B$, we have the expansion as a series in the parameter $\hbar$ shown in Figure \ref{fig:FusedBraiding-c}.
Now, in order to prove the identity \ref{fig:FusedReid3}, note that any morphism will always commute with the transposition $P_{A,B}$. Therefore, it suffices to prove the following commutation relation which is presented in diagrammatic form in Figure \ref{fig:ProofYB}.
The final step is to show that the last two diagrams in the figure cancel each other out. But this is evident as their orientations are reversed. To see this, note that the morphism $\alpha_T$, given by
\begin{equation}
    \alpha_T ~:~ T \otimes T \rightarrow T[1],
\end{equation}
can be shifted by $[-2]$ on both sides \cite{MR2661534}, giving
\begin{equation} \label{eq:L1Atiyah}
    \alpha_T ~:~ T[-1] \otimes T[-1] \rightarrow T[-1],
\end{equation}
which is now anti-symmetric and hence leads to the claimed cancellation. Indeed it is this version of the morphism which we have to use in order to evaluate Feynman diagrams as the $\chi$-fields are fermions. What we have just shown is the flatness condition \eqref{eq:ABflat}. This finishes the proof of the Yang-Baxter identity up to order $\hbar^2$. Sub-leading terms can be dealt with analogously.

\subsection{Invertible and non-invertible objects}

Let now $X$ be a K3 surface. We next want to compute $\mathrm{Ext}^2(A\otimes B, A \otimes B)$ for $A$ and $B$ two line bundles of the form $A=\mathcal{O}_X(C_1)$ and $B=\mathcal{O}_X(C_2)$ with $C_1$ and $C_2$ two curves in the K3. Thus, the task is to compute
\begin{equation}
    \mathrm{Ext}^2_X(\mathcal{O}_X(C_1) \otimes \mathcal{O}_X(C_2), \mathcal{O}_X(C_1) \otimes \mathcal{O}_X(C_2))
\end{equation}
Using 
\begin{equation}
    \mathrm{Ext}^n_X(\mathcal{E},\mathcal{F}) = H^n(X,\mathcal{E}^{\vee} \otimes \mathcal{F}),
\end{equation}
we thus have to compute
\begin{equation}
    H^2(X, \mathcal{E}^{\vee}\otimes \mathcal{E}),
\end{equation}
where $\mathcal{E} = \mathcal{O}_X(C_1) \otimes \mathcal{O}_X(C_2)$. But we know 
\begin{equation}
    \mathcal{E}^{\vee} = \mathcal{O}_X(-C_1)\otimes \mathcal{O}_X(-C_2),
\end{equation}
and hence we get
\begin{equation}
    H^2(X,\mathcal{E}^{\vee}\otimes \mathcal{E}) = H^2(X,\mathcal{O}_X) = \mathbb{C},
\end{equation}
generated by the holomorphic two-form of the K3 surface. This also shows that line bundles, viewed as objects in the category $\mathcal{C}$, are \textit{invertible}.

Let us next do the computation for a general vector bundle $E$ of rank $r+1$. That is, we want to compute
\begin{eqnarray}
    \mathrm{Ext}^2_X(E,E) = H^2(X,E \otimes E^{\vee}).
\end{eqnarray}
The rank $E \otimes E^{\vee}$ is simply $\mathrm{rk}(E)^2 = (r+1)^2$. Using that $c_1(E) = c_1(\mathrm{det}E) = c_1(L)$ for $L=\mathrm{det}E$ and $c_1(E^\vee) = c_1(\left(\mathrm{det}E\right)^\vee) =c_1(L^\vee) = -c_1(L)$, we see that from the properties of the Chern character we get
\begin{eqnarray}
    \mathrm{ch}(E \otimes E^\vee) & = & \mathrm{ch}(E) \mathrm{ch}(E^\vee) \nonumber \\
    ~ & = & \left(r+1 + c_1(E) + \mathrm{ch}_2(E) + \ldots \right)(r+1 - c_1(E) + \mathrm{ch}_2(E^\vee) + \ldots) \nonumber \\
    ~ & = & (r+1)^2 - c_1(E)^2 + 2(r+1) \mathrm{ch}_2(E) + \ldots 
\end{eqnarray}
Comparing this with 
\begin{equation}
    \mathrm{ch}(E \otimes E^\vee) = (r+1)^2 + c_1(E\otimes E^\vee) + \frac{c_1(E \otimes E^\vee)^2 - 2 c_2(E \otimes E^\vee)}{2} + \ldots
\end{equation}
and noting that $c_1(E \otimes E^\vee) = 0$, wee see that
\begin{equation}
    c_2(E \otimes E^{\vee}) = - 2 (r+1)\mathrm{ch}_2(E) + c_1(E)^2.
\end{equation}
Using $c_1(E)^2 = 2g - 2$ and $c_2(E) = \mathrm{deg}Z = d$, this gives 
\begin{equation}
   c_2(E \otimes E^\vee) = -r (2g-2)+2(r+1)d.
\end{equation}
Together with the Hirzebruch-Riemann-Roch formula for the Euler number of a coherent sheaf $\mathcal{F}$ on a K3 surface, 
\begin{equation}
    \chi(\mathcal{F}) = \frac{c_1(\mathcal{F})^2 - 2 c_2(\mathcal{F})}{2} + 2 \mathrm{rk}(\mathcal{F}),
\end{equation}
we finally obtain
\begin{equation} \label{eq:EE}
    \chi(E \otimes E^\vee) = 2 h^0(X,E \otimes E^\vee) - h^1(X,E \otimes E^\vee) = 2 - 2(g - (r+1)(r-d+g)).
\end{equation}
As a test of this formula, let us assume that $E$ is a line bundle $L$. Then $r=0$ and $d=0$ as the second Chern class vanishes. In this case our formula gives
\begin{equation}
    \chi(L \otimes L^\vee) = \chi(\mathcal{O}_X) = 2 - 2 (g - (0-0+g))=2,
\end{equation}
which agrees with the fact that on a K3 surface $h^0(X,\mathcal{O}_X) = h^2(X,\mathcal{O}_X) = 1$ and $h^1(X,\mathcal{O}_X)=0$ giving $\chi(X,\mathcal{O}_X) = h^0(X,\mathcal{O}_X) - h^1(x,\mathcal{O}_X) + h^2(X,\mathcal{O}_X) =2$. As a further example, let us consider the tangent bundle $E = \mathcal{T}_X$. Then $E^\vee = \mathcal{T}_X^\vee \equiv \mathcal{T}_X$ and moreover $H^0(X,S^m \mathcal{T}_X) = 0$. We compute
\begin{equation} \label{eq:ToT}
    E \otimes E^\vee = \mathcal{T}_X \otimes \mathcal{T}_X \equiv S^2 \mathcal{T}_X \oplus \mathcal{O}_X,
\end{equation}
and hence $h^0(X,\mathcal{T}_X \otimes \mathcal{T}_X) = h^2(X,\mathcal{T}_X \otimes \mathcal{T}_X) = 1$. Also, we see that $\mathcal{T}_X$ is a \textit{non-invertible} object in the fusion category. As for $E=\mathcal{T}_X$, $r=1$, $g=1$ (as $c_1(\mathcal{T}_X) = 0$) and $d = c_2(\mathcal{T}_X) = 24$, formula \eqref{eq:EE} gives
\begin{equation}
    \chi(\mathcal{T}_X \otimes \mathcal{T}_X) = 2 - 2 (1 - 2 (1-24+1))= - 88.
\end{equation}
This finally gives $h^1(X,\mathcal{T}_X \otimes \mathcal{T}_X) = 90$.

All these examples show that RW-theory is very special as compared to other 3d TQFTs in the following sense. Namely, even the Hilbert space corresponding to the two-point function on the sphere has higher dimension than $1$. That is there can be non-trivial morphisms (which are graded by even and odd cohomology in RW-theory) between two objects in the fusion category as shown in Figure \ref{fig:linejunction}.

\paragraph{a) Action of $H_{A,B}$ for $E_i = \mathcal{O}(C_i)$.}
Let us first consider the case that all bundles $E_i$ in \eqref{eq:HilbS2} are line bundles corresponding to some curves $C_i$, i.e. where $E_i = \mathcal{O}(C_i)$. In this case we have
\begin{equation}
    H^0(X, \mathcal{O}(C_1) \otimes \mathcal{O}(C_2) \otimes \cdots \otimes \mathcal{O}(C_m)) = H^0(X, \mathcal{O}(\sum_i C_i)).
\end{equation}
In case of algebraic K3 surfaces embedded in some projective space, the individual $\mathcal{O}(C_i)$ can be viewed as restrictions of $\mathcal{O}(n_i)$ of degree $n_i$ line bundles, and the tensor product will amount to
\begin{equation}
    H^0(X, \mathcal{O}(n_1) \otimes \mathcal{O}(n_2) \otimes \cdots \otimes \mathcal{O}(n_m)) = H^0(X, \mathcal{O}(N)),
\end{equation}
with $N= \sum_i n_i$. In this case, we can use Serre duality to compute the second cohomology,
\begin{eqnarray}
    H^2(X, \mathcal{O}(N)) & = & H^{2-2}(X, \mathcal{O}(N)^* \otimes \omega_X)^* \nonumber \\
    ~ & = & H^0(X, \mathcal{O}(-N) \otimes \omega_X)^* \nonumber \\
    ~ & = & H^0(X, \mathcal{O}(-N))^* = 0.
\end{eqnarray}
This shows that the action $H_{A,B}$ must be trivial unless $N=0$ in which case we have 
\begin{equation} \label{eq:Haction}
    H_{A,B} : H^i(X, \mathcal{O}) \mapsto \left\{\begin{array}{cl}
         H^{i+2}(X, \mathcal{O}), & \mathrm{for}~~ i=0\\
         0 & \mathrm{otherwise.} 
    \end{array}\right.
\end{equation}
The exponentiated map \eqref{eq:braiding} then gives
\begin{equation} \label{eq:tauaction}
    R_{A,B} : H^i(X,\mathcal{O}) \mapsto \left\{\begin{array}{cl}
         H^i(X,\mathcal{O}) \oplus \frac{\hbar}{2}H^{i+2}(X, \mathcal{O}), & \mathrm{for}~~ i=0\\
         H^i(X,\mathcal{O}), & \mathrm{otherwise.} 
    \end{array}\right.
\end{equation}
which is of upper triangular form.

\paragraph{b) Action of $H_{A,B}$ for $E_i = \mathcal{T}_X$.}
We now turn to the case where all bundles $E_i$ are equal to the holomorphic tangent bundle $\mathcal{T}_X$. In this case we can use identity \eqref{eq:ToT} to deduce that the Hilbert is nonempty if and only if we have an even number of such bundles in which case we have
\begin{equation}
    H^0(X,\mathcal{T}_X \otimes \cdots \otimes \mathcal{T}_X) = H^0(X,\mathcal{O}_X).
\end{equation}
Thus, via the action of $H_{A,B}$ and $R_{A,B}$, we obtain again the nilpotent transformation \eqref{eq:Haction} and the upper triangular transformation \eqref{eq:tauaction}.

\paragraph{c) Other Sheaves.}
More generally, one can construct sheaves of the form
\begin{equation}
    \mathcal{E}= \mathcal{F}_1 \oplus \mathcal{F}_2,
\end{equation}
where $\mathcal{F}_1$ and $\mathcal{F}_2$ are arbitrary sheaves. Such structures are called \textit{non-spherical} and one can expect a non-trivial Ext-algebra with contributions from
\begin{equation}
    \mathrm{Ext}^2(\mathcal{F}_i, \mathcal{F}_i), \quad \mathrm{Ext}^2(\mathcal{F}_i,\mathcal{F}_j), \quad i \neq j.
\end{equation}
This should make the map between $H^0(S, \bigotimes_l \mathcal{E}_l)$ and $H^2(S, \bigotimes_l \mathcal{E}_l)$ more interesting, though one should keep in mind that the map will be still nil-potent as the Ext-algebra truncates on K3:
\begin{equation}
    \mathrm{Ext}^i(\mathcal{E},\mathcal{E}) \circ \mathrm{Ext}^j(\mathcal{E},\mathcal{E}) = \mathrm{Ext}^{i+j}(\mathcal{E},\mathcal{E}), \quad \mathrm{Ext}^i = 0 \textrm{ for } i > 2.
\end{equation}

\subsection{Topological spins and R-matrix}

We can define topological spins via the action of $T$-transformation on the Hilbert space of a torus $T^2$. Whereas the state space specifies the representation of the objects in the category under the action of the Lie algebra, the simple objects themselves, also denoted as Anyons, are characterized by their image in the Grothendieck group $K(X)$ given by the Mukai vector as detailed in section \ref{sec:KX}. Similar to Chern-Simons theory, in the torus case, the Hilbert space can be written as the direct sum of Anyon types $\alpha$, such that\footnote{We assume here that the Grothendieck group is torsion free.}:
\begin{equation}
    \mathcal{H}(T^2) = \bigoplus_{\alpha} \mathbb{Z} v_{\alpha},
\end{equation}
where $v_{\alpha} \in K(X)$ denotes a generator in the Grothendieck group.
Let us next look into the action of the $SL(2,\mathbb{Z})$ generator $T$ on this Hilbert space, given by $\theta_{\alpha}$ which are exponentials of the topoligical spins defined in \eqref{eq:topspin},
\begin{equation} \label{eq:abTaction}
 T: v_{\alpha}  \mapsto \theta_{\alpha} v_{\alpha}.
\end{equation}
In the case of Rozansky-Witten theory this representation gets modified slightly and was worked out in the original paper \cite{Rozansky:1996bq}. We again specify to $X=\mathrm{K3}$. It was argued in \cite{Rozansky:1996bq} that, given the Hodge diamond of K3,
\begin{equation}
  \begin{array}{ccccccc}
    ~ & ~ & h^{0,0} & ~ & ~ & ~ \\
    ~ & h^{1,0} & ~ & h^{0,1} & ~ & ~ \\
    h^{2,0} & ~ & h^{1,1} & ~ & h^{0,2}\\
     ~ & h^{2,1} & ~ & h^{1,2} & ~ & ~ \\
     ~ & ~ & h^{2,2} & ~ & ~ & ~ 
  \end{array}
    = \quad 
  \begin{array}{ccccccc}
    ~ & ~ & 1 & ~ & ~ & ~ \\
    ~ & 0 & ~ & 0 & ~ & ~ \\
    1 & ~ & 20 & ~ & 1\\
     ~ & 0 & ~ & 0 & ~ & ~ \\
     ~ & ~ & 1 & ~ & ~ & ~ 
  \end{array},
\end{equation}
the action of $S$ and $T$ are trivial on the $H^{1,1}$ cohomology classes and are otherwise given by\footnote{Compared to the original reference \cite{Rozansky:1996bq}, we have restored the dependence on $\hbar$ and have swapped the roles of $H^{0,m}$ and $H^{2,m}$, which is possible as these spaces are dual to each other.}
\begin{eqnarray}
    \hat{S} & : & \begin{array}{ccl} 
                    H^{0,m} & \mapsto & H^{2,m} \\
                    H^{2,m} & \mapsto & H^{0,m}\\
                  \end{array},\\
    \hat{T} & : & \begin{array}{ccl}
                    H^{0,m} & \mapsto & H^{0,m} \oplus \frac{\hbar}{2} H^{2,m} \\
                    H^{2,m} & \mapsto &H^{2,m} \label{eq:TonH}\\
                  \end{array}.
\end{eqnarray}
Looking at the T-transformation, we see that it is block-diagonal with each block acting as the upper triangular matrix found in \eqref{eq:tauaction}. This is the analogue of \eqref{eq:abTaction} for the Rozansky-Witten MTC! Namely, instead of multiplication by topological spins $\theta_{\alpha}$ we have an action by the upper triangular matrix \eqref{eq:abTaction}. From this we learn two things. First, there are exactly 24 Anyons in the MTC, 20 of which have trivial representation and the remaining 4 split into two two-dimensional representations consisting of $\alpha=m=0$ and $\alpha=m=2$. Second, we have
\begin{equation}
    \theta \equiv \theta_i = \theta_j, \quad i \neq j
\end{equation}
and thus the R-matrix \eqref{eq:topspin} reduces to 
\begin{equation}
    R_{i,j} = \left(\theta_i \theta_j\right)^{-1} \theta_{i+j} = \theta^{-1},
\end{equation}
whose action is just the inverse of the action of \eqref{eq:tauaction} which at the level of cohomology is equivalent to the one of \eqref{eq:TonH}.

\subsection{Autoequivalences and non-invertible $0$-form symmetries}
\label{sec:KX}

In order to study autoequivalences of the derived category of K3, we first have to study its \textit{Grothendiek group} $K(X)$. This is the group by whose elements the simple objects in our category are labeled. It is the free abelian group generated by coherent sheaves $F$ on $X$ modded out by the subgroup which is generated by $[F_2] - [F_1] - [F_3]$ for short exact sequences of the form $0 \rightarrow F_1 \rightarrow F_2 \rightarrow F_3 \rightarrow 0$. The resulting space is then spanned by elements of the form $\sum n_i [F_i]$ with $n_i \in \mathbb{Z}$ and $F_i \in \mathrm{Coh}(X)$. If we pass over to the derived category $D^b(X) = D^b(\mathrm{Coh}(X))$, then there is the isomorphism 
\begin{equation}
    K(X) = K(\mathrm{Coh}(X)) \cong K(D^b(X)). 
\end{equation}
Any coherent free sheaf on $X$ admits a locally free resolution $0 \rightarrow F_n \rightarrow \ldots \rightarrow F_0 \rightarrow F \rightarrow 0$ and is thus represented as $[F] = \sum (-1)^i [F_i]$. When working with locally free sheaves, the tensor product induces on $K(X)$ the structure of a commutative ring by
\begin{equation}
    [F] \cdot [F'] \equiv [F \otimes F']. 
\end{equation}
For $X$ a K3 surface over an algebraically closed field $k$, elements of the Grothendieck group can be isomorphically represneted by their Chern character and one has the ring homomorphism
\begin{equation}
    \mathrm{ch} : K(X) \rightarrow \mathrm{CH}^*(X),
\end{equation}
where $\mathrm{CH}^*(X)$ is known as the \textit{Chow} group, and the Chern character of a sheaf $F$ on $X$ is given by
\begin{equation}
    \mathrm{ch}(F) = \mathrm{rk}(F) + c_1(F) + \frac{(c_1^2 - 2 c_2)(F)}{2}. 
\end{equation}
In the above, $\mathrm{rk}(F)$ is the rank of the sheaf $F$ and $c_1(F) = c_1(\det(F))$. Then, for a K3 surface over an algebraically closed field, we have the following correspondence
\begin{equation}
    \mathrm{CH}^0(X) \cong \mathbb{Z}, \quad \mathrm{CH}^1(X) \cong \mathrm{Pic}(X) \cong \mathbb{Z}^{\rho(X)},
\end{equation}
where $\rho(X)$ is the Picard number of $X$ and $\mathrm{CH}^2(X)$ is torsion free.

In order to see how Autoequivalences act on the cohomology of a complex K3 surface, we proceed as follows. First, on the integral cohomology
\begin{equation}
    H^*(X,\mathbb{Z}) = H^0(X,\mathbb{Z}) \oplus H^2(X,\mathbb{Z}) \oplus H^4(X,\mathbb{Z}),
\end{equation}
we introduce the \textit{Mukai pairing}
\begin{equation}
    \langle \alpha , \beta \rangle \equiv (\alpha_2 \cdot \beta_2) - (\alpha_0 \cdot \beta_4) - (\alpha_4 \cdot \beta_0),
\end{equation}
where $(~ \cdot ~)$ denotes the usual intersection pairing on $H^*(X,\mathbb{Z})$. $H^*(X,\mathbb{Z})$ with this pairing has signature $(4,20)$. This induces a weight-two Hodge structure defined by
\begin{equation}
    \widetilde{H}^{1,1}(X) \equiv H^{1,1}(X) \oplus (H^0 \oplus H^4)(X) \quad \textrm{and} \quad \widetilde{H}^{2,0}(X) \equiv H^{2,0}(X),
\end{equation}
determined by the condition that $\widetilde{H}^{(2,0)}(X)$ and $\widetilde{H}^{(1,1)}(X)$ are orthogonal with respect to the Mukai pairing. The numerical Grothendieck group can then be identified as 
\begin{equation}
    N(X) \cong \widetilde{H}^{1,1}(X) \cap \widetilde{H}(X,\mathbb{Z}) = (H^{1,1}(X) \cap H^2(X,\mathbb{Z})) \oplus (H^0 \oplus H^4)(X,\mathbb{Z}),
\end{equation}
such that the Mukai vector $v(E)$ of any $E \in D^b(X)$ can be written as 
\begin{equation}
    v(E) = (\mathrm{rk}(E),c_1(E),c_1(E)^2/2 - c_2(E) + \mathrm{rk}(E)) \in N(X) \subset \widetilde{H}(X,\mathbb{Z}). 
\end{equation}
Given a choice of a holomorphic two-form $\Omega \in H^{2,0}(X)$, the group $N(X)$ can be equivalently viewed as 
\begin{equation}
   N(X) = H^*(X,\mathbb{Z}) \cap \Omega^{\perp} \subset H^*(X,\mathbb{C}), 
\end{equation}
and has signature $(2,\rho(X))$. Then one has the following result. Namely, there is a group homomorphism
\begin{equation}
    \varrho: \mathrm{Aut}(\mathrm{D}^b(X)) \rightarrow \mathrm{Aut}(\widetilde{H}(X,\mathbb{Z})), \quad \Phi \mapsto \Phi^H,
\end{equation}
whose image is the subgroup $\mathrm{Aut}^+(\widetilde{H}(X,\mathbb{Z})) \subset \mathrm{Aut}(\widetilde{H}(X,\mathbb{Z}))$ of all orientation preserving Hodge isometries of $\widetilde{H}(X,\mathbb{Z})$. Examples of such isometries include
\begin{itemize}
    \item A \textit{spherical twist} $T_E : D^b(X) \stackrel{\sim}{\rightarrow} D^b(X)$ induces the following reflection on the cohomology
    \begin{equation}
        T^H_E = s_{v(E)} : \alpha \mapsto \alpha + \langle \alpha,v(E)\rangle \cdot v(E)
    \end{equation}
    in the hyperplane orthogonal to the $(-2)$-class $v(E) \in N(X)$.
    \item The autoequivalence of $D^b(X)$ given by $E \mapsto L \otimes E$ for some line bundle $L$ acts on cohomology via multiplication with $\exp(l) = (1,l,l^2/2) \in \widetilde{H}^{1,1}(X,\mathbb{Z})$, where $l = c_1(L)$.
\end{itemize}
The kernel of the map $\varrho$ is denoted by $\mathrm{Aut}^0(D^b(X))$. The functor $T_E^2$ defines a non-trivial element of $\mathrm{Aut}^0(D^b(X))$. Thus $T_E$ can be viewed as a \textit{non-invertible} $0$-form symmetry. By a conjecture of Bridgeland, the group $\mathrm{Aut}^0(D^b(X))$ is related to the fundamental group of a certain connected subset of the period domain.

\subsection{Derived categories and MTCs - a dictionary}

In this section we will collect some general characterization properties of the derived category of coherent sheaves on the one hand, and those of MTCs on the other, see table below.

\begin{table}[h]
\centering
\begin{tabular}{|c||c|c|}
\hline
~       & MTC            & $\mathrm{D}^b(X)$ \\
\hline
Objects & Simple Objects & Sheaves \\ 
Moduli & Davydov-Yetter & Complex Structure Deformations \\
Braiding & R-matrix & $\mathrm{Hom}_{D(X)}(A \otimes B, B \otimes A)$\\
Hilbert space & Conformal Blocks & Solutions of KZ-equation\\
Basis change & F-matrix & Associator\\
Automorphism & Galois Group & Fourier-Mukai twist\\
Representations & Quantum Group/Affine Algebra & Atiyah bracket \\
\hline
\end{tabular}
\caption{Derived categories and MTCs.}
\label{tab:DCMTC}
\end{table}

\section{RW-theory on K3 from Little String Theories (LSTs)}
\label{sec:LST}

According to \cite{Intriligator_2000} compactifying Little String Theories on a three-torus gives rise to a 3d $\mathcal{N}=4$ theory which on its Coulomb branch is described by a sigma model with Hyperk\"ahler target space. The corresponding target space can be $T^4$, $K3$ or instanton moduli space on these. A topological twist of such theories naturally leads to Rozansky-Witten theory. Let us therefore review the construction of LSTs from F-theory in the following.

\subsection{LSTs from F-theory}

Little String Theories can be geometrically engineered via F-theory by compactifying on a Calabi-Yau three-fold which is an elliptic fibration over a two complex-dimensinal non-compact K\"ahler surface $\mathcal{S}$ \cite{Bhardwaj_2016}. The 6d theory gives rise to strings from D3 branes wrapping compact curves in the base, where the string charge lattice of the theory is identified with the lattice $\Lambda = H_2(\mathcal{S},\mathbb{Z})$. We have a natural quadratic pairing given by the intersection form
\begin{equation}
    (\cdot, \cdot) : H_2(\mathcal{S},\mathbb{Z}) \times H_2(\mathcal{S},\mathcal{Z}) \rightarrow \mathbb{Z}~.
\end{equation}
This intersection form is negative definite in the case of 6d SCFTs and semi-negative definite for LSTs. The latter criterion ensures that not all compact curves in the base are shrinkable and there is a unique homology class of self-intersection zero which sets the LST scale. Denoting the non-shrinkable curve of the LST by $\Sigma^0$ and the other curves by $\Sigma^I$, the negative of the intersection form is 
\begin{equation}
    \eta^{IJ} = - (\Sigma^I, \Sigma^J) \quad I,J = 1,\ldots, r+1,
\end{equation}
where the integer $r$ is the dimension of the tensor branch of the theory. The D3 branes are charged under the gauge group $g_I$ coupled with the $I$-th tensormultiplet.

\paragraph{Defect group.} The 2-form part of the defect group of the theory is given by
\begin{equation}
    \mathbb{D}^{(2)} = \mathbb{Z} \oplus \bigoplus_{j=1}^p \IZ_{m_j},
\end{equation}
where the integers $m_j > 1$ are the diagonal entries of the Smith normal form of $\eta^{IJ}$. The first factor of the defect group corresponds to the the LS charge arising from a D3 brane wrapping the non-shrinkable curve in the base.

\subsection{RW-theory from compactification}

As stated above, compactification of the LST on $T^3$ leads to a sigma model with Hyperk\"ahler target. The dimension of the target space $X$ is given by the formula
\begin{equation} \label{eq:6dmoduli}
    \dim_{\IR}X = 4 (r_V + n_T),
\end{equation}
where $r_V$ is the total rank of the 6d gauge group while $n_T$ is the number of tensor multiplets. The effective 3d sigma model has then $\mathcal{N}=4$ supersymmetry, where the supercharges transform as $(\mathbf{2},\mathbf{2})$ under $SU(2)_E \times SU(2)_R$. Here, $SU(2)_E$ denotes the Euclidean rotation group in three dimensions and $SU(2)_R$ is the R-symmetry group. One can twist the theory by taking the new rotation group $SU(2)'$ to be the diagonal subgroup of $SU(2)_E \times SU(2)_R$. Then the supercharges transform as two copies of $\mathbf 1 \oplus \mathbf 3$ under $SU(2)'$ and in particular there are two $SU(2)'$-invariant supercharges. This then leads to the topological field theory of Rozansky and Witten. Note that the fermions of the twisted theory also transform as $\mathbf 1 \oplus \mathbf 3$ under $SU(2)'$ and hence correspond to a zero-form $\eta$ and a one-form $\chi_{\mu}$ with the following action:
\begin{eqnarray}
    S & = & \int_M L_0 + L_1, \nonumber \\
    L_0 & = & \half \Omega_{ij}\chi^i \nabla \chi^j - \frac{1}{6}R(\eta, \chi,\Omega \chi, \chi), \nonumber \\
    L_1 & = & d \phi \star d\bar \phi - \nabla \eta \star \chi. 
\end{eqnarray}
In the above $\phi$, $\bar \phi$ are complex coordinates of $X$ and we have the map
\begin{equation}
    \phi: W_3 \rightarrow X,
\end{equation}
where $W_3$ denotes the three-dimensional spacetime of the 6d LST perpendicular to $T^3$. The fermionic fields $\chi$ are sections of $\Omega^1_M \otimes \phi^* T^{1,0} X$ and moreover $\eta \in \phi^* T^{0,1}X$.

\subsection{Hyperk\"ahler geometry}

We next would like to discuss the Hyperk\"ahler geometry arising from $T^3$-compactified little string theory. An expression for the corresponding metric at the classical level (i.e. without quantum corrections) was already derived in \cite{Intriligator_2000}
\begin{equation}
    ds^2 = \frac{\sqrt{\det h}}{g_6^2} (h^{-1})^{ab} d\phi_a d\phi_b + \frac{g_6^2}{\sqrt{\det h}}(d\phi_4 - \theta^a d\phi_a)^2, \quad a=1,2,3, 
\end{equation}
where $g_6$ is the gauge coupling of the six-dimensional theory\footnote{Which in our case is a tensor multiplet expectation value.} and the three scalars $\phi_a$ arise from Wilson lines of the gauge field around the cycles of $T^3$. Moreover, the 3d $U(1)$ gauge field has been dualized to the scalar $\phi_4$. The particular little string theory we would like to focus on is the theory describing $N$ small, coincident $SO(32)$ instantons in the heterotic string which leads to a 6d LST with $\mathcal{N}=(1,0)$ SUSY. This leads at low energies to an $Sp(N)$ gauge group which for $N=1$ is just $SU(2)$. As argued in \cite{Intriligator_2000}, due to the $\mathbb{Z}_2$ Weyl action of $SU(2)$, the low-energy Coulomb branch of this theory when compactified on a three-torus is a K3 surface. 

Following \cite{Poon}, we introduce a different notation by redefining
\begin{equation}
    a(u) \equiv \phi_1 + i \phi_2
\end{equation}
and
\begin{equation}
    \theta_e \equiv \phi_3, \quad \theta_m \equiv \phi_4, \quad z \equiv \frac{\theta_m - \tau \theta_e}{2\pi}.
\end{equation}
We take the total geometry to be an elliptically fibered K3 surface
\begin{equation} \label{eq:fiber}
    y^2 = x^3 + f(u,w) x + g(u,w),
\end{equation}
such that $[u,w]$ are projective coordinates on the base $\IP^1$ of the fibration and \eqref{eq:fiber} defines the elliptic fiber with complex structure parameter $\tau$ determined by the $j$-function
\begin{equation}
    j(\tau) \equiv \half \frac{(24f)^3}{4 f^3 + 27 g^2},
\end{equation}
which has $24$ singular fibers at points $u_a$ (have set $w=1$) where the discriminant 
\begin{equation}
    \Delta = 4 f^3 + 27 g^2
\end{equation}
vanishes. In the above, $f$ and $g$ are homogeneous polynomials with degrees $8$ and $12$. Then we have
\begin{equation}
    da = \eta^2 \left[\prod_{a=1}^{24} (u-u_a)^{-1/12}\right]du, 
\end{equation}
and using the fiberwise differential $\tilde{d}z$ which treats $u$ as constant:
\begin{equation}
    \tilde{d}z = \frac{d\theta_m-\tau d\theta_2}{2\pi}.
\end{equation}
Using these differentials, the holomorphic 2-form is given by
\begin{equation}
    \omega_+ = da \wedge dz = da \wedge \tilde z,
\end{equation}
and the K\"ahler form becomes
\begin{equation}
    \omega_3 = \frac{i}{2} \left(R \tau_2 da \wedge d\bar{a} + \frac{1}{R\tau_2} \tilde{d}z \wedge \tilde{d}\bar{z}\right).
\end{equation}

\subsection{Global symmetries}

We would like to put our focus on 1-form continuous global $U(1)$ symmetries which are present in 6d gauge theories and are associated to the 2-form instanton current,
\begin{equation} \label{eq:instdef}
    J^{(2)} \sim \star \mathrm{Tr}\left(f^{(2)} \wedge f^{(2)}\right).
\end{equation}
Such operators have been recently analyzed for 6d theories in \cite{Cordova:2020tij} and we will be following that reference for our exposition here.
Integration of the hodge dual of this 2-form current over a four-dimensional surface $\Sigma_4$ defines then a conserved charge operator,
\begin{equation}
    Q(\Sigma_4) = -i \int_{\Sigma_4} \star J^{(2)} .
\end{equation}
The defects carrying $Q$ charge are line defects and are linked with the surface $\Sigma_4$. Acting with such a codimension $2$ operator on the vacuum gives rise to a string defect that is charged under $Q$ which in our case is quantized, i.e. $Q \in \mathbb{Z}$. The current $J^{(2)}$ can be coupled to a 2-form background gauge field $B^{(2)}$, giving rise to an additional term in the action of the form
\begin{equation} \label{eq:bjcoupling}
    \int B^{(2)} \wedge \star J^{(2)},
\end{equation}
and conservation of the current requires invariance under background gauge transformations $B^{(2)} \rightarrow B^{(2)} + d\Lambda^{(1)}$. Note that in 6d SCFTs the 2-form field $B^{(2)}$ is gauged and therefore the coupling \eqref{eq:bjcoupling} implies that $J^{(2)}$ is trivial as a global symmetry current \cite{Cordova:2020tij}. However, this is not the case in little string theories and therefore here the continuous 1-form symmetry corresponding to line defects is preserved! 

Apart from the usual 1-form symmetries, the little string theory has also more traditional 0-form symmetries. Moreover, the fusion of $0$-form symmetries into a $2$-form current of the form discussed above leads to so-called continuous $2$-group global symmetry, to which we turn next. Denote by $j^{(1)}_G$ the conserved current of a continuous $0$-form flavor symmetry $G^{(0)}$. Apart from such flavor symmetries there is always the $0$-form symmetry associated to Poincar\'e symmetry $\mathcal{P}^{(0)}$. Thus we obtain the following picture for the global symmetries of our 6d little string theory,
\begin{equation}
    G^{(0)} \times \mathcal{P}^{(0)} \times U(1)^{(1)}.
\end{equation}
When $2$-group symmetries are present, the background gauge transformations for $B^{(2)}$ are modified and there will be also shifts under $G^{(0)}$ background gauge symmetries, which we parametrize by $\lambda_G^{(0)}$, and a contribution from local Lorentz transformations denoted by $\theta^{(0)a}_b$,
\begin{equation} \label{eq:mixedanomalies}
    B^{(2)} \rightarrow B^{(2)} + d\Lambda^{(1)} + \frac{\hat{\kappa}_G}{4\pi} \mathrm{Tr}\left(\lambda^{(0)}_G d A^{(1)}_G\right) + \frac{\hat{\kappa}_{\mathcal{P}}}{16 \pi}\mathrm{tr}\left(\theta^{(0)} d\omega^{(1)}\right),
\end{equation}
where $\hat{\kappa}_G$, $\hat{\kappa}_{\mathcal{P}}$ are known as $2$-group structure constants, and $\omega^{(1)a}_b$ is the spin connection of the Riemannian background geometry. One then denotes the $2$-group global symmetry by writing
\begin{equation}
    \left(G^{(0)} \times \mathcal{P}^{(0)}\right) \times_{\hat{\kappa}G,\hat{\kappa}_{\mathcal{P}}} U(1)^{(1)}.
\end{equation}
Consistency requires that the structure constants are quantized ,
\begin{equation}
    \hat{\kappa}_G \in \mathbb{Z}, \quad \hat{\kappa}_{\mathcal{P}} \in \mathbb{Z}.
\end{equation}
In the case of the little string theory arising from small heterotic $SO(32)$ instantons, the global flavor symmetry is $G^{(0)} = SU(2)_L^{(0)} \times SU(2)_R^{(0)} \times SO(32)^{(0)}$ where the factor $SU(2)_R^{(0)}$ is an R-symmetry. If we denote the gauge group by $g^{(0)} = Sp(N)$, the authors of \cite{Cordova:2020tij} determine the corresponding $2$-group structure constants to be
\begin{eqnarray}
    \hat{\kappa}_L = -(N-1), & \quad &\hat{\kappa}_R = h^{\vee}_g, \nonumber \\
    \hat{\kappa}_{SO(32)} = -1, & \quad & \hat{\kappa}_{\mathcal{P}} = 2.
\end{eqnarray}

\subsection{Defect moduli space}

Since the line defect $\mathcal{L}$ in six dimensional gauge theories (with gauge group $G$) is an instanton configuration as described in \eqref{eq:instdef}, it admits a representation in terms of D0 branes in LSTs arising from NS5 branes in type IIA string theory. The worldvolume of these ``heavy'' D0 branes (their mass scales as the inverse of the string coupling which is sent to zero), is then the line defect in question. To study the corresponding moduli space, one has to study instanton solutions on 4-cycles which link this line defect. In our case, since 3 directions are already compactified on a $T^3$ and the defect is extended along a line within the perpendicular 3-manifold $M_3$, the only 4-cycle which links this line is composed of a circle linking $\mathcal{L}$ within $M_3$ together with $T^3$. Thus we see that we need $\Sigma_4 \equiv T^4$ and therefore the moduli space $\mathcal{M}_{\mathcal{L}}$ of our line defect is the moduli space of $G$-instantons on $T^4$. 

Remarkably, one can identify the Coulomb branch moduli space of the LST compactified on $T^3$ also with instanton moduli spaces on $T^4$ (or K3 in more general situations), where we refer to \cite{Haghighat:2018dwe} for a recent discussion. One way to see this correspondence is the case of $N$ M-theory (or heterotic) five-branes probing a $\mathbb{C}^2/\Gamma_G$ singularity where $G$ is of ADE type. Both situations can be represented in terms of M5 branes probing $\mathbb{C}^2/\Gamma_G \times S^1$ (respectively $\mathbb{C}^2/\Gamma_G \times S^1/\mathbb{Z}_2$). Compactifying three dimensions of the five-brane on a $T^3$, one can first take one of them to be the M-theory circle and then successively perform two T-dualities along the other two, giving $N$ D2 branes. Uplifting back to M-theory, these become $N$ M2 branes probing $\mathbb{C}^2/\Gamma_G \times T^4$ (respectively $\mathbb{C}^2/\Gamma_G  \times \mathrm{K3}$). Since the M2 branes are localized at the origin of the singularity, one sees that they are small $G$-instantons on $T^4$ (respectively K3) \cite{Intriligator_2000}. To count the dimension of such instanton moduli spaces we employ the Atiyah-Hitchin-Singer index theorem \cite{Atiyah} giving
\begin{equation} \label{eq:instmoduli}
    \mathrm{dim}_{\mathbb{R}}\mathcal{I}_{k,G} = 4 h_G^{\vee} k - \frac{1}{2}\mathrm{dim}G(\chi - \sigma) = 4 h_G^{\vee} k,
\end{equation}
where $h_G^{\vee}$ is the dual Coxeter number of the gauge group $G$, and $\chi$ and $\sigma$ are Euler number and signature of the four-manifold in question respectively. In our case, we have used that $\chi(T^4) = \sigma(T^4) = 0$. One can then convince oneself that this matches the one of the Coulomb branch moduli space given in \eqref{eq:6dmoduli}.

\subsection{Identification with autoequivalences of $D^b(K3)$}

In the following we would like to see how the above generators of $0$-form symmetries in six dimensions act as autoequivalences of the derived category of K3 upon compactification on $T^3$. In other words, we expect a rank-preserving embedding
\begin{equation}
    \varphi : \left(G^{(0)} \times \mathcal{P}^{(0)}\right)_{T^3} \hookrightarrow \mathrm{Aut}(D^b(K3)).
\end{equation}
This can be understood as follows. Recall that for $X=K3$, the lattice $H^2(X,\mathbb{Z})$ is an even unimodular lattice of signature $(3,19)$ which is isomorphic to the lattice 
\begin{equation}
   \Lambda_{\mathrm{K3}} \equiv E_8(-1)^{\oplus 2} \oplus U^{\oplus 3}, 
\end{equation}
where $U$ is the hyperbolic lattice 
\begin{equation}
    U = \left(\begin{array}{cc} 0 & 1 \\ 1 & 0 \end{array}\right),
\end{equation}
and $E_8(-1)$ is the negative of the Cartan matrix of $E_8$. Since we have two copies of the $E_8$ Cartan matrix, we have exactly sixteen $-2$ curves which can be used to spherically twist as discussed in Section \ref{sec:symmetries}. We identify these with the $16$ Cartan generators of the $SO(32)$ global symmetry. Upon compactification on $T^3$ these generators will act on the Coulomb branch of the resulting 3d IR theory which is precisely our K3 surface. As the spherical twist $T_E^2$ is an element of $\mathrm{Aut}^0(D^b(K3))$, we see that all these $16$ reflections actually form \textit{non-invertible} walls in our 3d theory. From a purely 3d field theoretical perspective, we can understand these non-invertible walls as interfaces where on one side a $\mathbb{Z}_2$ $0$-form symmetry  corresponding to a Weyl reflection is gauged simulaneously with a $\mathbb{Z}_2$ $1$-form symmetry (see for example \cite{Cui:2024cav} for a recent discussion). This is possible if the two global symmetries have no mixed anomalies \eqref{eq:mixedanomalies} which happens precisely for the $16$ Cartan generators.

What is left to understand is the action of group $\mathcal{P}^{(0)}$. Here, note that upon compactification of $T^3$, the Poincare symmetry is broken and now contains a discrete subgroup which is the mapping class group of $T^3$. This group is $SL(3,\mathbb{Z})$ and has two order $3$ and one order $2$ generator. Using these generators one can form three independent order $2$ generators corresponding to swapping of two circles and a reflection on the third. We identify these generators with the action of the three $U$-generators of the K3 cohomology which at the same time are Hodge isometries.

\section{Applications}
\label{sec:applications}

In this section we explore a number of applications of the ideas developed in this paper. Among them are compactifications of 4d $\mathcal{N}=2$ theories on $S^1$, Affine Grassmannians, as well as reductions of 6d $(2,0)$ theories on 3-manifolds $M_3$ and braiding in the resulting theory $T[M_3]$.

\subsection{Wild ramification and 4d $\CN=2$ theories on $S^1$}

Part of the motivation for the present paper was to generalize categories of line operators $\CC = \text{MTC} [S^1 \times \Sigma]$ in topologically twisted class-S theories compactified on a circle to more general RW theories, {\it cf.} \cite{Dedushenko:2018bpp}. In particular, when $\Sigma$ is a genus-0 curve with wild ramification, the corresponding 4d $\CN=2$ theory is an Argyres-Douglas SCFT. The simplest example in this family is the so-called $(A_1, A_2)$ theory. It has the following gravitational anomaly coefficients
\be
a = \frac{43}{120} \qquad , \qquad c = \frac{11}{30}
\ee
and after compactification on $S^1$ gives rise to a topological sigma-model with target space $X$, a non-compact K3 manifold corresponding to an elliptic fibration with a singular fiber of Kodaira type II.

In this example, the corresponding category of topological line operators
\be
\CC = \text{MTC} [S^1 \times \Sigma] = D(X)
\label{4dMTC}
\ee
depends on an additional parameter (a root of unity of order 5) which recovers the entire Galois orbit of Lee-Yang MTC and Fibonacci MTC. All these MTCs have two simple objects, and
\be
K^0 (\CC) = \mathbb{Z} \oplus \mathbb{Z}
\ee
in agreement with the fact that a resolution of $X$ leads to a manifold with $H^* (\tilde X) = \mathbb{Z} \oplus \mathbb{Z}$.

The generalization to other examples is straightforward. For instance, the Argyres-Douglas theory $(A_1, A_3)$ has $a=\frac{11}{24}$, $c = \frac{1}{2}$ and $X$ an elliptic fibration of Kodaira type III. It has $H^* (\tilde X)$ of rank 3, in perfect agreement with the fact that the corresponding MTC \eqref{4dMTC} has three simple objects \cite{Dedushenko:2018bpp}. Instead of continuing with generalizations, let us return to the $(A_1, A_2)$ theory and take a closer look at its 0-form symmetries and the spectrum of foams (topological surface operators). It is shown in \cite{Gukov:2016lki,Fredrickson:2017yka} that this theory has indeed a $\mathbb{Z}_5$ global symmetry. In order to see this symmetry in terms of a corresponding monodromy in moduli space as envisioned in Section \ref{sec:symmetries}, we employ mirror symmetry which in the case of non-compact K3 surfaces just gives the complex structure deformation of the Kodaira singularity \cite{Katz:1997eq}. The monodromy will then show up as a $\mathbb{Z}_5$ monodromy of K3 periods which according to a conjecture of \cite{Argyres:2016yzz} is arising from a Kodaira type $II^*$ singularity (see table 1 there). Thus we have the following correspondence,  
\be
\text{Auteq} \, D^b (\textrm{Kodaira}(II)) = \mathbb{Z}_5.
\ee
Moreover, the previously mentioned $(A_1,A_3)$ theory arising from the type III Kodaira singularity flows in the IR upon gauging a discrete $0$-form symmetry to a theory classified by a type $IV^*$ Kodaira singularity with no flavor symmetry but a discrete $\mathbb{Z}_3$ 2-form symmetry (see again reference \cite{Argyres:2016yzz}). Now, employing again mirror symmetry, the monodromy of such type $IV^*$ singularities was explicitly worked out in \cite{Lerche:1998nx} where the result was shown to be $\mathbb{Z}_3$. Thus we conclude that 
\be
    \text{Auteq} \, D^b (\textrm{Kodaira}(III)) = \mathbb{Z}_3.
\ee
This also matches with the Galois symmetry of the original $(A_1,A_3)$ theory.

\subsection{Affine Grassmannian}

Note, all above examples are self-mirror (in the sense of homological mirror symmetry, i.e. 2d mirror symmetry). Therefore, instead of $D(X)$ one can consider the suitable (derived) version of the Fukaya category $\CF (X)$. It turns out that it is the latter that generalizes to $\text{MTC} [M_3]$ for more general 3-manifolds.

Indeed, for general 3-manifolds, not of the form $M_3 = S^1 \times \Sigma$, one can not identify $\text{MTC} [M_3]$ with $D(X)$ because general theories $T[M_3]$ only have 3d $\CN=2$ supersymmetry. And, it was argued in \cite{Gukov:2016gkn} that $\text{MTC} [M_3]$ is basically the Fukaya-Seidel category of the space of all $G_{\mathbb{C}}$-connections on $M_3$ with the superpotential given by the Chern-Simons functional. This category is very close to the Fukaya category of $\CM_{\text{flat}} (G_{\mathbb{C}}, M_3)$. Note, the latter is a (singular) K\"ahler space for general $M_3$. The category $\text{MTC} [M_3]$ can be thought of as the category of line operators in 3d $\CN=2$ theory $T[M_3]$, similar to the category of line operators in Rozansky-Witten theory reviewed in section~\ref{sec:symmetries}.

Thus, in a sigma-model whose target space $X$ is the total space of the cotangent bundle to the (affine) Grassmannian, the algebra of lines / walls / interfaces on the one hand is the group algebra of $\text{Auteq} \, D^b (X)$ and, on the other hand, is given by the equivariant K-theory of the corresponding Steinberg variety, equipped with the convolution product. This algebra is the affine (resp. double affine) Hecke algebra, more precisely its spherical part.

The case of affine Grassmannian is especially interesting since it provides a good case study for infinite-dimensional $X$. Since in such cases one does not have a finite quiver realization of $X$, it is generally harder to identify the space of FI parameters (stability parameters) relevant to \eqref{Gparam}. The case of affine Grassmannian is also interesting because in this case the role of Steinberg is played by the space of triples $\CR$ introduced in \cite{Braverman:2016pwk,Braverman:2016wma}. Its equivariant K-theory gives (quantized) K-theoretic Coulomb branch. In particular, in the classical limit the spectrum of the equivariant K-theory is $\left( \mathbb{T} \times \mathbb{T}^{\vee} \right) / W$, in agreement with the decorated structure (i.e. generalized / categorical symmetries) of the corresponding TQFT \cite{Jagadale:2022abr}, which is expected to be the TQFT associated with quantum groups $U_q (\frak{g})$ at generic $q$.

These considerations naturally lead us to the following equivalence of the (suitably defined) derived categories
$$
D^b (\CR) \; \simeq \; sDAHA\text{-mod} \simeq \;
U_q^{\cdots} (\frak{g})\text{-mod}^{\cdots}
$$
where the category on the right-hand side should include infinite-dimensional Verma modules.
This conjecture can be compared to the realization --- due to Arkhipov, Bezrukavnikov, and Ginzburg \cite{MR2053952} --- of the derived category of the principal block in modules over the Lusztig (``big'') quantum group $U$ as the derived category of equivariant coherent sheaves on the Springer resolution\footnote{Note, the derived category of the principal block $D^b \text{block}^{\text{mix}} (U)$ is sometimes denoted $D^b \text{Rep}_{\oslash} (G)$.}
$$
D^b \text{block}^{\text{mix}} (U) \; \cong \; D^b \text{Coh}^{G \times \CC^*} (\tilde \CN) \; \cong \; D^b \text{Perv}^{\text{mix}} (\text{Gr}_G)
$$
Omitting the ``equivariant'' part (i.e. performing the de-equivariantization) gives a similar equivalence for the small quantum group, {\it cf.} \cite{MR2349618}:
\be
u_q (\frak g)-\text{mod}^0 \; \cong \; \text{Coh} (\tilde \CN)
\ee

\subsection{Reduction of $(2,0)$ theory on 3-manifolds}
\label{sec:MTCM3}

Compactification of an M5 brane on a 3-manifold $M_3$ obtained through plumbing characterized by a graph $\upsilon$ leads to \cite{Gadde:2013sca} $U(1)^n$ $\mathcal{N}=2$ Chern-Simons action which, using the three-dimensional $\mathcal{N}=2$ vector superfield $V = (A_{\mu},\lambda,\sigma,D)$ and the field strength $\Sigma = \overline{D}^{\alpha} D_{\alpha} V$, is given by 
\begin{equation} \label{eq:3daction}
	\mathcal{L} = \sum_{i,j=1}^n \int d^3 x d^4 \theta \frac{K_{ij}}{4\pi} V_i \Sigma_j, 
\end{equation}
where $K$ is the linking matrix of the plumbing graph and $n = \textrm{rank}(K)$. To obtain the corresponding bosonic action, one uses
\begin{equation}
	S = \frac{1}{4\pi} \int d^3 x d^4 \theta V_i \Sigma_j = \frac{1}{4\pi} \int (A \wedge dA - \bar{\lambda}\lambda + 2 D \sigma).
\end{equation}
We call the theory defined by this action $T[M_3]$. The bosonic part of the action \eqref{eq:3daction}, given by
\begin{equation} \label{eq:baction}
	\mathcal{L}_{\textrm{bosonic}} = \sum_{i,j=1}^n \frac{K_{ij}}{4\pi}\int A_i \wedge d A_j 
\end{equation}
defines then a TQFT $\mathrm{MTC}[M_3]$ which is characterized by the group $H_1(M_3;\IZ)$ as follows, where we will assume that $b_1 = 0$, i.e. $H_1(M_3;\IZ)$ consists only of torsion classes, for now. Consider $\Lambda_K$ to be the integral lattice $\IZ^n$ in $n$-dimensional space $\IR^n$, spanned by basis vectors $\nu_1,\ldots,\nu_n$ which we take to form the rows of a matrix $M$. Then a generic vector $x = (x_1,\ldots,x_n) \in \Lambda_Q$ can be written as 
\begin{equation}
	x = \xi_1 \nu_1 + \cdots + \xi_n \nu_n = \xi M,
\end{equation}
where $\xi_i$ are integers and $\xi = (\xi_1,\ldots,\xi_n)$. One can take $M$ to be the identity matrix to simplify matters in which case $x = (\xi_1,\ldots,\xi_n)$. The lattice is characterized by the fact that the norm of the vector $x$ is computed in terms of the quadratic form $K$:
\begin{equation} \label{eq:qform}
	\langle x,x \rangle = \sum_{i=1}^n \sum_{j=1}^n \xi_i \xi_j ~\nu_i ~K~ \nu_j^t = \xi ~K~ \xi^t. 
\end{equation}
Quantization of the Chern-Simons action \eqref{eq:baction} on the geometry $T^2 \times \IR_+$ then leads to a quantum Hilbert space $\mathcal{H}_{T^2}$ on the boundary given by
\begin{equation}
	\Phi_{\vec{\alpha}} |0\rangle , \quad \vec{\alpha} \in H_1(M_3;\mathbb{Z}) = \Lambda_K^* / \Lambda_K,
\end{equation} 
where $|0\rangle$ is a unique vacuum state. In case the determinant of $K$ is non-vanishing, the dual lattice can be characterized by
\begin{equation}
    \Lambda^*_K = \{ K^{-1} \xi ~|~ \xi \in \IZ^n\},
\end{equation}
and the quadratic form on $\mathscr{D} \equiv \Lambda^*_K/\Lambda_K$ becomes $\mathbb{Q}/\IZ$-valued.
The boundary theory can be viewed as a chiral vertex operator algebra which can be identified as the left-moving part of a 2d $\mathcal{N}=(0,2)$ theory \cite{Feigin:2018bkf}. In this correspondence the chiral vertex operators can be identified with the $\Phi_{\vec{\alpha}}$ and the operator product becomes the group law of the lattice:
\begin{equation} \label{eq:latticefusion}
	\Phi_{\vec{\alpha}} \cdot \Phi_{\vec{\beta}} \sim \Phi_{\vec{\alpha}+\vec{\beta}}.
\end{equation}

\paragraph{$b_1 > 0$.}

Here we should describe what happens to $\text{MTC} [M_3]$ when $b_1 (M_3) > 0$. To understand this more general situation, it is instructive to first review the well-known situation where $K$ is invertible \cite{Belov:2005ze,Delmastro:2019vnj}. In this situation the theory contains Anyon or vortex worldlines in the bulk which are labelled by elements $\vec{\alpha}$ of the discriminant group $\mathscr{D}$ as defined above. These Wilson lines are of the form 
\begin{equation}
    W_{\vec{\alpha}}(\gamma) \equiv \exp\left[i \vec{\alpha}^t \int_{\gamma} A \right],
\end{equation}
where $A$ is the $U(1)^n$ gauge field $A = (A_1,A_2,\ldots,A_n)^t$. $\vec{\alpha}$ can be viewed as the spin of particles with fractional statistics. Wavefunctions corresponding to boundary states of Abelian Chern-Simons theory with such Anyon worldlines have been constructed in \cite{Belov:2005ze}. Such wavefunctions can be seen as conformal blocks and are theta functions which can be interpreted as sections of line bundles over an Abelian variety. 

Here one subdivides between two situations. When all diagonal components of $K$ are even, then it follows that all local operators of theory are bosonic and one speaks of a bosonic TQFT. In this case there are $|\det K|$ independent sections of the line bundle corresponding to $|\det K|$ independent line operators. Here $|\det K|$ is the number of lattice points in the discriminant domain $\mathscr{D}$ which is bounded by an $n$-dimensional parallelepiped. However, when one or more diagonal components of $K$ are odd, the theory contains fermionic degrees of freedom and requires a choice of spin structure. In this case one speaks of a spin TQFT. Here line operators of the theory are labelled by points in the lattice $\mathscr{D} \times \mathbb{Z}_2$ as lines can differ by a local fermion even if congruent modulo $K$. Thus one has $2 |\det K|$ independent lines.

In order to extract the Anyon fusion rules of an Abelian Chern-Simons theory, or equivalently its one-form symmetry group, one can utilize the construction of wavefunctions in terms of theta functions. The corresponding theta functions can be seen to be equivalent if two different Chern-Simons level matrices $K$ and $\widetilde{K}$ have the same Smith normal form. That is, if there exist matrices $P, Q \in SL(n,\IZ)$ and $\widetilde{P}, \widetilde{Q} \in SL(n,\IZ)$ such that
\begin{equation}
    P^t K Q = \widetilde{P}^t \widetilde{K} \widetilde{Q} = D,
\end{equation}
where $D$ is the diagonal matrix
\begin{equation}
    D = \left(\begin{array}{ccc}d_1 & ~ & ~\\ ~ & \ddots & ~\\ ~ & ~ & d_n\end{array}\right),
\end{equation}
with all elements $d_1,\ldots,d_n \in \IZ$. In this case the discriminant group is given by
\begin{equation}
    \mathscr{D} = \bigoplus_{i=1}^n \IZ_{d_i}.
\end{equation}
In case $b_1 > 0$ for our 3-manifold $M_3$, then some of the diagonal entries of $D$ are zero and we have
\begin{equation}
    D = \left(\begin{array}{cccc}d_1 & ~ & ~ & ~\\ ~ & \ddots & ~ & ~\\ ~ & ~ & d_{n-m} & ~ \\ ~ & ~ & ~ & \mathbf{0}_{m\times m}\end{array}\right), \quad \Rightarrow \quad \mathscr{D} = \bigoplus_{i=1}^{n-m} \IZ_{d_i} \oplus \IZ^m.
\end{equation}
The discriminant group $\mathscr{D}$ can then be identified with the group of 1-form symmetries of the TQFT.

\subsection{Braiding in $T[M_3]$ and lessons for $\text{MTC} [M_3]$}

While $\mathcal{N}=4$ theories in three dimensions admit full topological twist (in fact, more than one), three-dimensional theories with only $\mathcal{N}=2$ supersymmetry can not be made fully topological as they only have a $U(1)$ R-symmetry. Such a theory can not be put on an arbitrary 3-manifold $W_3$ as the Lorentz symmetry on a general $W_3$ is $SO(3)$ which can not combine with $U(1)_R$ to produce a topological twist. In order to be able to perform a topological twist we put $T[M_3]$ on geometries of the form 
\begin{equation}
    W_3 = \IR_t \times \Sigma,
\end{equation}
with $\Sigma$ an arbitrary Riemann surface. In this case, the Lorentz symmetry breaks down to $U(1)_{\Sigma}$ and one can perform a topological twist by the diagonal embedding
\begin{equation}
    U(1)'_{\Sigma} = U(1)_{\Sigma} \times U(1)_R.
\end{equation}
It has been argued in \cite{Gukov:2016gkn} that such theories associated to 3-manifolds via 3d-3d correspondence via a partial topological twist produce an MTC-like algebraic structure that encodes anomalies and symmetries of $T[M_3, G]$ theories:
\begin{equation}
\text{MTC} [M_3 , G] = \text{anomalies / symmetries}
\end{equation}
So far, detailed understanding of this algebraic structure has been mainly limited to the modular data (see e.g. \cite{Dedushenko:2018bpp,Gukov:2020lqm,Gukov:2022cxv} for explicit computations) and the main question of this paper --- whether it admits braiding and, if so, of what kind --- is very natural. This question is one of the main motivations for the present work.

Therefore, in order to see how this important question might be answered, let us take a closer look at the definition of $\text{MTC} [M_3 , G]$ and compare it to the derived category of coherent sheaves underlying the Rozansky-Witten twist of 3d $\mathcal{N}=4$ theories. As we will see shortly, while the latter is purely geometric, the latter is more ``quantum'' in the sense that it is defined via quantum geometry, i.e. quantum K-theory in the A-model twist of 3d $\mathcal{N}=2$ theory $T[M_3, G]$. Indeed, this is basically the definition of $\text{MTC} [M_3 , G]$ according to \cite{Gukov:2016gkn}.

In the $\mathcal{N}=4$ case, simple objects in the category can be understood as fixed points of the $U(1)_t$ symmetry acting on the Coulomb branch, when the fixed points are isolated. The equivariant $U(1)_t$ localization on the Coulomb branch can be thought of as a way to regularize non-compactness of typical Coulomb branches in 3d $\mathcal{N}=4$ theories. Theories with $\mathcal{N}=2$ supersymmetry, on the other hand, do not possess an additional $U(1)_t$ symmetry that can be used in this way, but they also do not need such a regularization since quantum effects generate a twisted superpotential $W(x)$ on the Coulomb branch of a generic 3d $\mathcal{N}=2$ theory. The instances when this does not happen are precisely fine tuned 3d $\mathcal{N}=2$ theories where supersymmetry is enhanced to $\mathcal{N}=4$.

Therefore, the analogue of $U(1)_t$ fixed points that give simple objects in the category for $\mathcal{N}=4$ theory should be critical points of the twisted superpotential $W(x)$. When things are generic enough, such critical points are isolated and sit in bounded regions of 3d $\mathcal{N}=2$ moduli spaces (``Coulomb branches'').

For example, when $G=U(1)$ the theory $T[M_3, G]$ is free. On a circle, $S^1$, which is the origin of quantum K-theory, it can be described as a product of a sigma-model into $\IC \times \IC^*$ and a TQFT determined by the linking form on $H_1 (M_3)$. Schematically,
\begin{equation}
T [M_3, U(1)] = \IC \times \IC^* \times \text{TQFT}
\end{equation}

\begin{figure}[h!]
\centering	\includegraphics[width=0.8\textwidth]{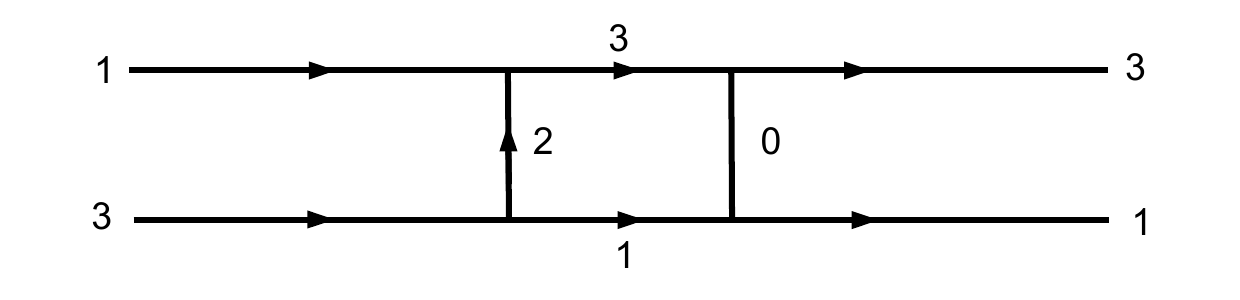}
  \caption{Implementation of braiding using topological symmetry along Riemann surface $\Sigma$.}
  \label{fig:Braiding}
\end{figure}

\begin{figure}[h]
\centering
\includegraphics[width=0.8\textwidth]{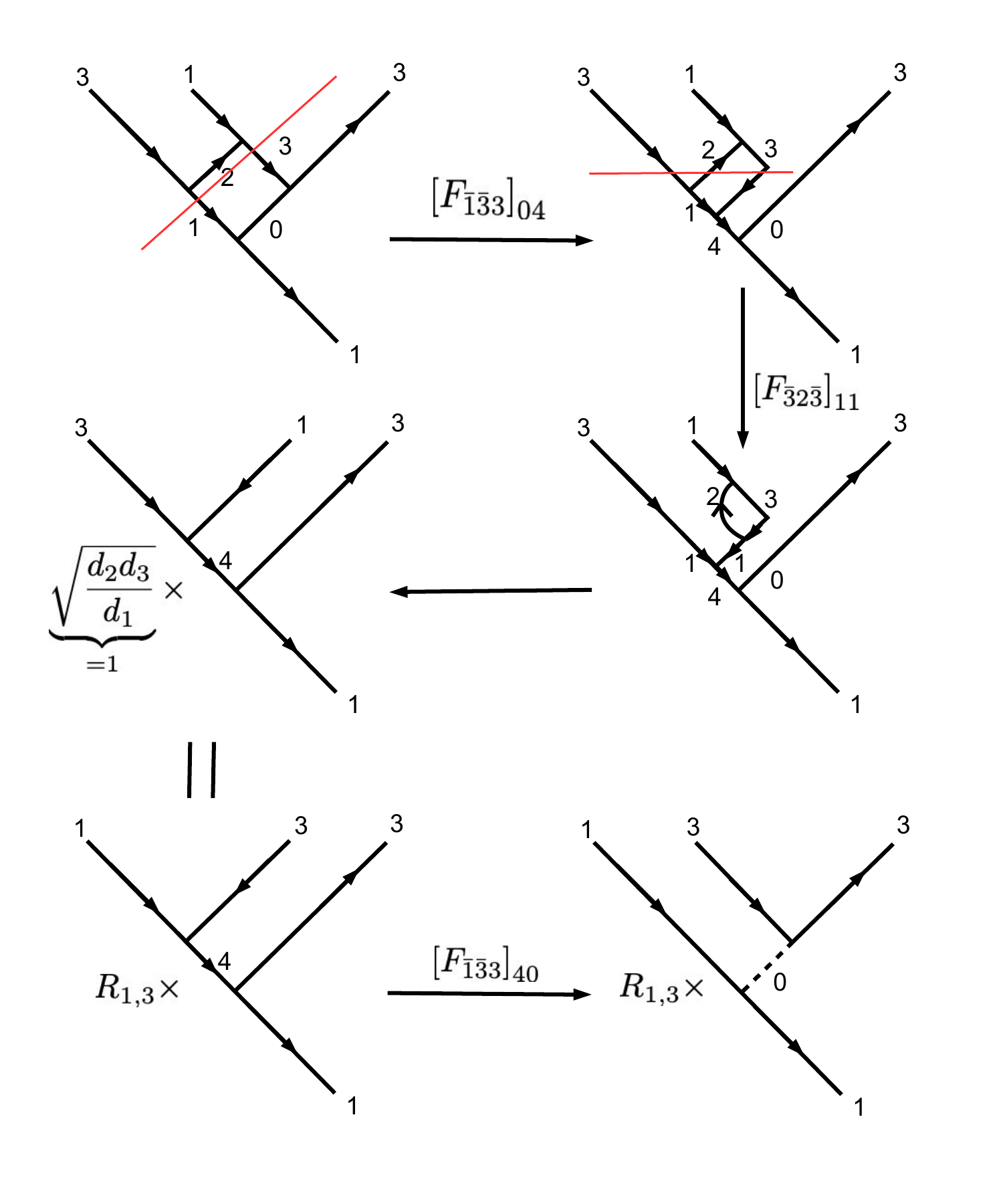}
  \caption{Evaluation of topological braiding using fusion diagram giving $R$-matrix. }
  \label{fig:REvaluation}
\end{figure}

In this context, we introduce the concept of \textit{networks} to facilitate braiding of line defects on $W_3$. For example, to implement the braiding between a type $A=1$ Anyon and a type $B=3$ Anyon, we use the diagram in Figure \ref{fig:Braiding}.
Braiding is implemented using the topological twist along the Riemann surface where line operators can bend and form networks.
The diagram is then evaluated using the fusion category rules \cite{Aasen:2020jwb} which in this case gives the identity operator multiplied by $R_{1,3}$, see Figure \ref{fig:REvaluation}.
Line operators in $T[M_3]$, as well as vacua, are labeled by $G_{\mathbb{C}}$ flat connections on $M_3$ where $G_{\mathbb{C}}$ depends on the number of M5 branes wrapping $M_3$. In the case of a single M5 brane as reviewed above, $G_{\mathbb{C}} = U(1)$ and the vacua of the theory are given by Abelian representations of the fundamental group $\pi_1(M_3)$ into $U(1)$, modulo conjugation, i.e. (see \cite{Gadde:2013sca})
\begin{equation}
    \mathcal{V}_{T[M_3;U(1)]} = \mathrm{Rep}(\pi_1(M_3) \rightarrow U(1))/\mathrm{conj.} = H_1(M_3).
\end{equation}
Thus our space of vacua merely consists of a collection of discrete points. This should be contrasted with the case in the RW-theory where the space of vacua in the little string compactification on $T^3$ is a Hyperk\"ahler manifold. Proceeding analogously to RW-theory, we then expect objects of our MTC to correspond to sheaves on the space of vacua. However, in this case, the only possible sheaves are skyscraper sheaves on the different points and thus we get an identification with our Anyons.

\acknowledgments{It is our pleasure to thank Alexander Braverman, Will Donovan, Boris Feigin, Daniel Huybrechts, Hiraku Nakajima, Justin Sawaon, Yan Soibelman, Johannes Walcher and Maxim Zabzine for helpful discussions and suggestions.
The work of S.G. is supported by a Simons Collaboration Grant on New Structures in Low-Dimensional Topology, by the NSF grant DMS-2245099, and by the U.S. Department of Energy, Office of Science, Office of High Energy Physics, under Award No. DE-SC0011632.} The work of B.H. is supported by NSFC grant 12250610187. B.H. would also like to thank the Max-Planck institute of Mathematics in Bonn, where part of this work was completed, for hospitality and financial support. The work of N.R. was supported by the Simons Collaboration ``Categorical symmetries'',  by the grant BMSTC and ACZSP (Grant no. Z221100002722017) and by the Changjiang fund.

\bibliographystyle{JHEP}     
 {\small{\bibliography{main}}}

\end{document}